\documentclass[astrosymb,resetfootnote,longbib]{aastex7}

\usepackage{xspace}
\usepackage{graphicx}

\definecolor{mylinkcolor}{RGB}{120, 0, 180} 

\shorttitle{14 Her c JWST/NIRCam}
\shortauthors{Bardalez Gagliuffi \& Balmer, et al.}

\hypersetup{linkcolor=mylinkcolor,citecolor=mylinkcolor,urlcolor=mylinkcolor}

\newcommand{\um}{\text{\,\textmu m}\xspace}
\newcommand{\teff}{$T_\mathrm{eff}$}
\newcommand{\logg}{log\,$g$}

\newcommand{\mjup}{$M_\mathrm{J}$}

\submitjournal{The Astrophysical Journal Letters}

\graphicspath{{./}{figures/}}

\begin{document}

\title{JWST Coronagraphic Images of 14 Her c: \\a Cold Giant Planet in a Dynamically Hot, Multi-planet System}

\author[0000-0001-8170-7072,sname=Bardalez Gagliuffi]{Daniella C. Bardalez Gagliuffi}
\altaffiliation{These authors contributed equally to this work}
\email[show]{dbardalezgagliuffi@amherst.edu}
\affiliation{Amherst College, Department of Physics and Astronomy, Department of Physics \& Astronomy, Amherst College, 25 East Drive, Amherst, MA 01002, USA}
\affiliation{American Museum of Natural History, Department of Astrophysics, 200 Central Park West, New York, NY 10024, USA}

\author[0000-0001-6396-8439,sname=Balmer]{William O. Balmer}
\altaffiliation{These authors contributed equally to this work}
\email[show]{wbalmer1@jhu.edu}
\affiliation{Department of Physics \& Astronomy, Johns Hopkins University, 3400 N. Charles Street, Baltimore, MD 21218, USA}
\affiliation{Space Telescope Science Institute, 3700 San Martin Drive, Baltimore, MD 21218, USA}

\correspondingauthor{Daniella C. Bardalez Gagliuffi \& William O. Balmer}

\author[0000-0003-3818-408X]{Laurent Pueyo}
\email{pueyo@stsci.edu}
\affiliation{Space Telescope Science Institute, 3700 San Martin Drive, Baltimore, MD 21218, USA}

\author[0000-0003-2630-8073]{Timothy D. Brandt}
\email{tbrandt@stsci.edu}
\affiliation{Space Telescope Science Institute, 3700 San Martin Drive, Baltimore, MD 21218, USA}

\author[0000-0002-0078-5288]{Mark R. Giovinazzi}
\email{mgiovinazzi@amherst.edu}
\affiliation{Amherst College, Department of Physics and Astronomy, Department of Physics \& Astronomy, Amherst College, 25 East Drive, Amherst, MA 01002, USA}

\author[0000-0003-3130-2282]{Sarah Millholland}
\email{sarah.millholland@mit.edu}
\affiliation{Department of Physics, Massachusetts Institute of Technology, Cambridge, MA 02139, USA}
\affiliation{Kavli Institute for Astrophysics and Space Research, Massachusetts Institute of Technology, Cambridge, MA 02139, USA}

\author{Brennen Black} 
\email{brennen@mit.edu}
\affiliation{Department of Physics, Massachusetts Institute of Technology, Cambridge, MA 02139, USA}

\author[0000-0003-0834-8645]{Tiger Lu} 
\email{tiger.lu@yale.edu}
\affiliation{Department of Astronomy, Yale University, New Haven, CT 06511, USA}

\author[0000-0002-7670-670X]{Malena Rice} 
\email{malena.rice@yale.edu}
\affiliation{Department of Astronomy, Yale University, New Haven, CT 06511, USA}

\author[0000-0001-5864-9599]{James Mang} 
\email{j_mang@utexas.edu}
\affiliation{Department of Astronomy, University of Texas at Austin, Austin, TX 78712, USA}

\author[0000-0002-4404-0456]{Caroline Morley} 
\email{cmorley@utexas.edu}
\affiliation{Department of Astronomy, University of Texas at Austin, Austin, TX 78712, USA}

\author[0000-0002-9420-4455]{Brianna Lacy} 
\email{brianna.i.lacy@gmail.com}
\affiliation{Department of Astronomy, University of California, Santa Cruz, CA 95064, USA}
\affiliation{Bay Area Environmental Research Institute + NASA Ames Research Center, Moffett Field, CA 94035, USA}

\author[0000-0001-8627-0404]{Julien Girard} 
\email{jgirard@stsci.edu}
\affiliation{Space Telescope Science Institute, 3700 San Martin Drive, Baltimore, MD 21218, USA}

\author[0000-0003-0593-1560]{Elisabeth Matthews} 
\email{matthews@mpia.de}
\affiliation{Max-Planck-Institut für Astronomie, D-69117 Heidelberg, Germany}

\author[0000-0001-5365-4815]{Aarynn Carter} 
\email{aacarter@stsci.edu}
\affiliation{Space Telescope Science Institute, 3700 San Martin Drive, Baltimore, MD 21218, USA}

\author[0000-0003-2649-2288]{Brendan P. Bowler} 
\email{bpbowler@ucsb.edu}
\affiliation{Department of Physics, University of California, Santa Barbara, CA 93106, USA}

\author[0000-0001-6251-0573]{Jacqueline K. Faherty} 
\email{jfaherty@amnh.org}
\affiliation{American Museum of Natural History, Department of Astrophysics, 200 Central Park West, New York, NY 10024, USA}

\author[0000-0002-2428-9932]{Clemence Fontanive} 
\email{clemence.fontanive@umontreal.ca}
\affiliation{Institute for Research on Exoplanets, University of Montr\'{e}al, Montr\'{e}al, QC H3A 2A7, Canada}

\author[0000-0003-4203-9715]{Emily Rickman} 
\email{erickman@stsci.edu}
\affiliation{European Space Agency (ESA), ESA Office, Space Telescope Science Institute, 3700 San Martin Drive, Baltimore MD, 21218}

\begin{abstract}

Most observed multi-planet systems are coplanar, in a dynamically ``cold" configuration of concentric orbits like our own Solar System. With the James Webb Space Telescope (JWST) we have detected 14 Her c, the first mature and cold exoplanet directly imaged in a dynamically ``hot", multi-planet system. With large eccentricities and a nonzero mutual inclination, the present-day architecture of this system points to a turbulent past and ongoing angular momentum exchange between the planetary orbits of 14 Her b and c. The temperature of 14 Her c rivals both the coldest imaged exoplanet and the coldest known brown dwarf. Moreover, its photometry at 4.4\um is consistent with the presence of carbon disequilibrium chemistry and water ice clouds in its atmosphere. 14 Her c presents a unique laboratory to study giant planet formation, dynamical evolution of multi-planet system architectures, and atmospheric composition and dynamics in extremely cold worlds.

\end{abstract}


\section{Introduction}

The early evolution of our own Solar System was dominated by the dynamical influence of the gas giants \citep{2024arXiv240414982R}. After the dispersal of gas in the protoplanetary disk, Jupiter may have scattered the outer giant planets onto wider orbits, and even possibly prompted the ejection of a fifth giant planet \citep{2011ApJ...742L..22N}. The final outcome of the migration of the giant planets was our stable Solar System, with Jupiter redirecting ice-rich small bodies inwards as one potential pathway to deliver water to the terrestrial planets and to allow life to thrive on Earth \citep{doi:10.1126/sciadv.adp2191}. The putative ejected giant planet fared the opposite fate, and was consigned to wander the Solar neighborhood without a host star as a ``rogue planet." Whether similar histories of dynamical interaction, planet scattering, and ejection are common across planetary systems remains an open question due to the difficulty of precisely mapping exoplanet system architectures and constraining the occurrence rate of isolated rogue planets \citep{Sumi2023}. In this paper, we describe the orbital architecture of a nearby exoplanetary system whose dynamically rich history has produced an architecture radically different from our Solar System's, with two giant planets on misaligned, eccentric orbits, potentially as a result of a planetary ejection. The study of such an extreme planetary system is enabled by our JWST \citep{Rigby2023} detection of the potentially coldest and oldest planet yet imaged.

14 Herculis is a nearby ($d = 17.898\pm0.009$\,pc; \citealt{2021AandA...649A...1G}), metal-rich ([Fe/H] = $0.43\pm0.07$; \citealt{2010AandA...515A.111S}), K0-type, solar mass  star hosting two giant planets \citep{2021ApJ...922L..43B}. These two massive planets cause a reflex motion on the star detectable as a periodic change in its radial velocity and proper motion. Radial velocity monitoring spanning over 20 years revealed a recurrent 5-year period due to the inner planet, 14 Her b, and a long-term trend~\citep{2007ApJ...654..625W} later confirmed as the widely-separated exoplanet 14 Her c by incorporating absolute astrometry from the Hipparcos and Gaia missions \citep{2021ApJ...922L..43B}. 
This analysis showed that the two planets in the system are roughly located at the equivalent semi-major axes of the asteroid belt and Neptune's orbits in our own Solar System ($a_b = 2.845^{+0.039}_{-
0.03}$\,au, $a_c = 27.4^{+7.9}_{-16}$\,au), yet on very eccentric ($e_b = 0.369\pm0.003$, $e_c = 0.64\pm0.13$), extremely misaligned orbits ($\Theta_{bc} = 96.3^{+36.8}_{-29.1}\,^{\circ}$). Only a handful of other multi-planet systems have a measured orbital misalignment between two of their planetary orbits (e.g., $\pi$ Men, \citealt{2020A&A...640A..73D,2020MNRAS.497.2096X}; $\upsilon$ And, \citealt{2010ApJ...715.1203M,2015ApJ...798...46D}; Kepler-108, \citealt{2017AJ....153...45M}; HAT-P-11, \citealt{2020MNRAS.497.2096X, An2025, Lu2025}; HD 73344, \citealt{2024arXiv240809614Z}; HD 3167, \citealt{2021A&A...654A.152B}), yet none of these planets have been directly imaged. While direct imaging is a challenging technique due to the large contrast between star and planet, it can provide firsthand information on the planet, its exact location, and its atmosphere. 

The measured dynamical masses of the 14 Her planets ($M_b = 9.1\pm1.9\,$\mjup, $M_c = 6.9^{+1.6}_{-1.0}\,$\mjup) at the age of the system ($4.6^{+3.8}_{-1.3}$\,Gyr) imply they have extremely cold temperatures ($250-300\,\mathrm{K}$;~\citealt{2021ApJ...922L..43B}), making them far more comparable to our own Jupiter (125\,K;~\citealt{1981JGR....86.8705H}) than the majority of directly imaged exoplanets \citep[700-1700\,K;][]{2016PASP..128j2001B}, except for 
$\epsilon$ Indi Bb~\citep{2024Natur.633..789M}, and TWA-7b~\citep{2025arXiv250215081L}. Cold exoplanets ($<500\,K$) emit most of their light in the mid-infrared, and are therefore difficult to image from the ground in the optical or near-infrared, because of their intrinsic faintness and large flux ratio, or `contrast,' with their host stars. With its space-based, mid-infrared capabilities, JWST has the unique ability to directly image much colder planets than have been studied before.

In this paper we present the first direct image of the nearby, cold, mature, giant planet 14~Her~c with JWST/NIRCam coronagraphy and revise its orbital parameters. This is the first direct image of a planet in a significantly misaligned planetary system. In Section~\ref{sec:obs} we describe the new direct imaging observations with JWST/NIRCam and the process to measure the photometry and astrometry of 14 Her c. In Section~\ref{sec:analysis_results} we describe the updated orbit fitting, evolutionary and atmospheric modeling, and dynamical simulations of the orbits. In Section~\ref{sec:discussion}, we discuss our results and their implications for the mutual inclination of the planetary orbits, the likelihood of disequilibrium chemistry in the planet's atmosphere, and the time evolution of the planetary orbits. We present our conclusions in Section~\ref{sec:conclusions}. Technical details about our methods on image reduction, orbit fitting, evolutionary and atmospheric modeling, and N-body simulations are summarized in the Appendix (Sections~\ref{sec:img_red}-\ref{sec:atm_mod}). 

\section{Observations and data reduction}~\label{sec:obs}

We used JWST/NIRCam \citep{Rieke2023} to image 14 Her c as the main science objective of Cycle 2 program GO 3337 (PI: Bardalez Gagliuffi, Co-PI: Balmer). Our observing strategy relied on both Angular Differential Imaging (ADI) and Reference Differential Imaging (RDI) to perform starlight subtraction \citep[][for a recent review]{Follette2023}. For validation, we used the \texttt{pyNRC}\footnote{https://pynrc.readthedocs.io/en/latest/} \citep{pynrc} and \texttt{PanCAKE}\footnote{https://aarynncarter.github.io/PanCAKE} Python packages to determine optimal instrument readout patterns and estimate the contrast performance of the observations. Our NIRCam observations of 14 Her c were carried out between May 18, 2024 18:31:45 and May 19, 2024 01:08:45 (UTC) for a total of 7.8 hours (including observatory overheads) simultaneously with the F200W and F444W filters and using the MASKA335R coronagraph \citep{Krist2010} at position angles $191.87^\circ$ and $181.87^\circ$. This science observation used the \texttt{SHALLOW4} readout pattern at 10 groups per integration and 96 integrations per exposure. Immediately following the observation of the target, we observed the reference star HD~144002 using the \texttt{SHALLOW4} readout pattern, 8 groups per integration, and 16 integrations per exposure, with the \texttt{9-POINT-CIRCLE} small grid dither pattern in an uninterruptable sequence to preserve the similarity of the telescope wavefront between observations. Observing the reference star with the small grid dither increases the diversity of the reference Point Spread Function (PSF) library \citep{Soummer2014}. We selected this reference star based on its relative on-sky proximity to 14~Her ($\rho{=}3.67^\circ$ in J2016), its brighter \textit{K}-band magnitude, yet similar spectral type, and its low \textit{Gaia} Renormalised Unit Weight Error \citep[RUWE = $0.986$,][]{Lindegren2018}, which indicates that it is likely a single star.

The observations were designed to detect a $250\,\mathrm{K}$ source at 1\farcs3 from the star at a contrast of $10^{-6}$ with a $5\,\sigma$ confidence in the F444W filter, while simultaneously collecting parallel images in the F200W filter where the planet signal was not expected. The limit in F200W-F444W color allows for the rejection of the vast majority of potential background contaminants, including stars, galaxies, and quasars, which would be at least as bright in F200W in comparison to F444W. The MASKA335R mask was chosen to balance inner working angle (IWA), raw starlight suppression performance, and reliable target acquisition. 

Image reduction and calibration followed previous work on the JWST/NIRCam coronagraph \citep{Carter2023,2024ApJ...974L..11F,Balmer2025b}, the details of which are recorded in Appendix~\ref{sec:img_red}\footnote{A notebook that can be used to reproduce the data reduction and image analysis described in Appendix~\ref{sec:img_red} can be found online at the DOI:~\doi{10.5281/zenodo.15483953} \citep{reduction_doi}}. The residual starlight in the coronagraphic images was modeled and removed using the Karhunen--Lo\`{e}ve Image Projection (KLIP) algorithm \citep{Soummer2012} through the Python implementation \texttt{pyKLIP}\footnote{https://pyklip.readthedocs.io/en/latest/} \citep{Wang2015} wrapped by \texttt{spaceKLIP}\footnote{https://spaceklip.readthedocs.io/en/latest/}. 
\par
\begin{figure*}
    \centering
    \includegraphics[width=\linewidth]{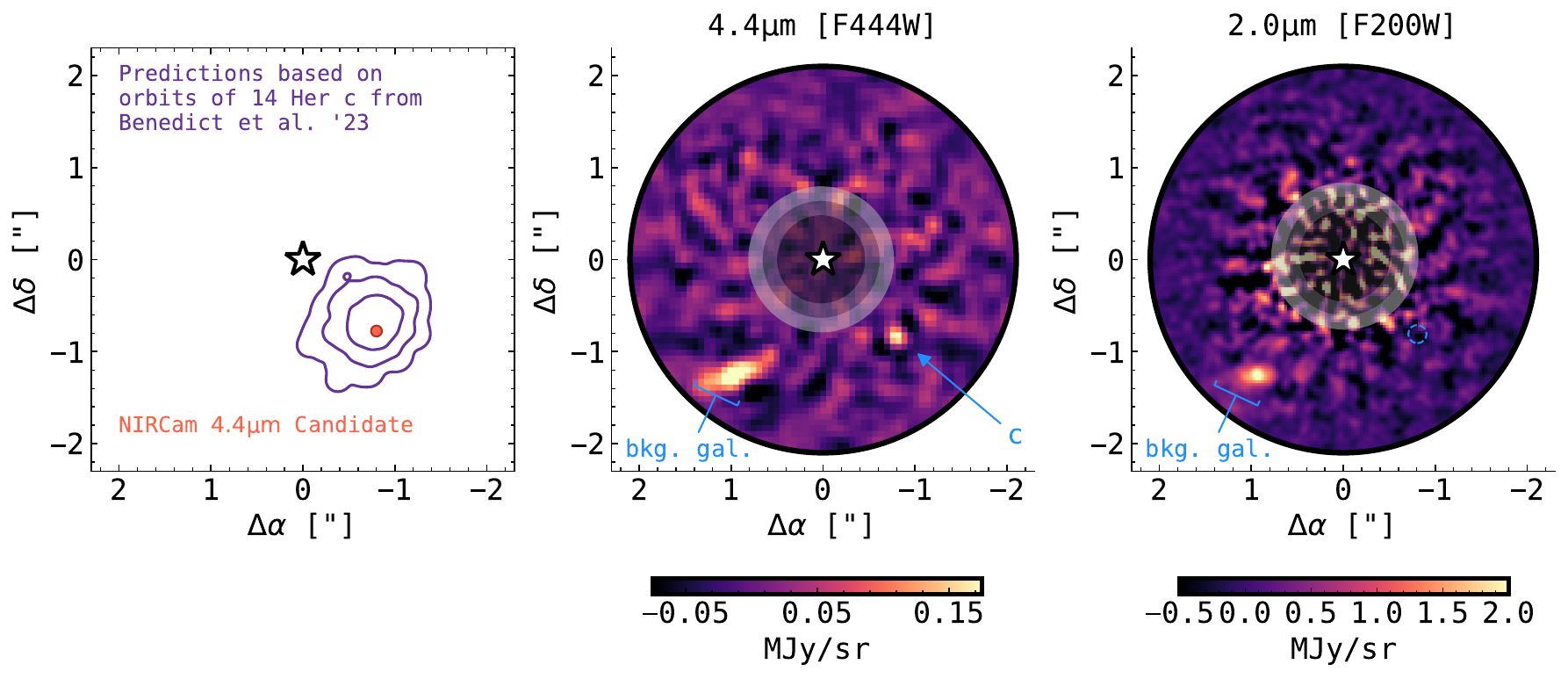}
    \caption{\textbf{First direct image of 14 Her c}. JWST/NIRCam coronagraphic imaging of the 14~Her system with the \texttt{MASKA335R} coronagraph (transmission indicated by gray shaded circles). North is up, east is left.
    \textit{Left:} Prediction for the location of 14~Her~c from a reproduction of the orbits presented in \citep{2023AJ....166...27B}, at the epoch of observation (MJD=60449), and the location of the point source detected in the F444W filter. \textit{Middle:} A starlight-subtracted image taken with the F200W filter ($1.755{-}2.227 \, \mathrm{\mu m}$), showing extended emission from a background galaxy at about 1\farcs5 to the SE from the host star. No other statistically significant point sources are apparent, as intended, given the cold temperature of the planet 14~Her~c. \textit{Right:} A starlight-subtracted image taken with the F444W filter ($3.881{-}4.982 \, \mathrm{\mu m}$), showing extended emission from the same background galaxy, and a point source at 1\farcs1 to the SW with a contrast of $1\times10^{-6}$. The location of the point source, detected with a contrast of $9.6\times10^{-7}$, agrees within $1\sigma$ with the predicted location for the planet, so we conclude we have detected 14~Her~c.}
    \label{fig:NIRCam_img}
\end{figure*}

 The JWST image of 14 Her c is shown in Figure~\ref{fig:NIRCam_img}. We detected a point source in the F444W filter with a signal-to-noise ratio of 5.7\,$\sigma$, an angular separation of $1\farcs115\pm0\farcs005$ at a position angle of $225.84^{\circ}\pm0.26^{\circ}$ East of North, and apparent magnitude of $19.68\pm0.07\,\mathrm{mag}$ with reference to Vega. There is also a spatially resolved background galaxy 1\farcs5 away and to the SE of the star in both F200W and F444W images, whereas the point source only appears in the F444W image as expected for a cold giant planet. The location of our point source is consistent within $1\,\sigma$ with the latest predictions of the location of 14~Her~c based on orbital solutions from the literature \citep[reproduced from their Table 10]{2023AJ....166...27B}. 

\subsection{Contrast curves}
 
\par We estimated the contrast performance of our observations using the \texttt{spaceKLIP} package following previous work \citep{Kammerer2022, Carter2023}. We measured the annular standard deviation of the starlight-subtracted images, masking the positions of the two sources within 2\farcs0 (the planetary point source and the bright extended galaxy). The raw contrast was corrected for small number statistics in the innermost annuli using the Student t-distribution correction following \citet{Mawet2014}. The corrected contrast was then calibrated into physical units using a synthetic \texttt{BT-NextGen} model stellar spectrum \citep{Allard2011} with parameters \teff${=}5300\,\mathrm{K}$, $\log(g){=}4.5$, and $\mathrm{[Fe/H]}{=}+0.5$ \citep{1999ApJ...511L.111G,2021ApJ...922L..43B}, scaled to match the Hipparcos/Tycho2 \citep{2000A&A...355L..27H}, Gaia \citep{GaiaCollaboration2022}, and 2MASS \citep{skr06} photometry reported for 14 Her. 
The coronagraph transmission function\footnote{The NIRCam \texttt{MASKA335R} coronagraph transmission is an axisymmetric function of separation from the center of the mask, see Figure 4 in \citet{Krist2010}.} was applied to the calibrated contrast. Finally, we determined the starlight subtraction throughput via injection-recovery tests\footnote{We refer the reader to \S3.1 and \S3.2 in \citet{Carter2023} for additional details.}. Figure \ref{fig:cc} shows our final calibrated contrast curves. The observations are sensitive to sources with contrast ${\sim}4{\times}10^{-7}$ at 1\farcs0 in the F444W filter, and ${\sim}3{\times}10^{-7}$ at 1\farcs0 in the F200W filter at $5\,\sigma$ confidence, slightly exceeding our anticipated contrast performance of $10^{-6}$ at 1\farcs0.

\subsection{Photometry and astrometry of 14~Her~c}

\par The process of the starlight subtraction with ADI and RDI oversubtracts astrophysical sources present in the image. We forward-modeled this degradation of the PSF \citep{Pueyo2016} in order to obtain accurate and unbiased astrometry and photometry of 14~Her~c using \texttt{pyklip} \citep{Wang2015, Golomb2021} and \texttt{spaceKLIP} \citep{Kammerer2022, Carter2023, Kammerer2024, Franson2024}. Briefly, an off-axis coronagraphic PSF model was generated using \texttt{webbpsf} at the approximate position of the candidate. The model was scaled, translated, and subtracted from the data to compute a likelihood function, while the scale and translation parameters were sampled iteratively using the \texttt{MultiNest} (via \texttt{pyMultiNest}) sampling algorithm \citep{Feroz2009, Buchner2014} with 400 live points. We accounted for the presence of correlated noise in the image by sampling the length-scale of a Gaussian process, parameterized using a Mat\'ern $\nu=3/2$ kernel, which approximates residual speckle noise \citep{Wang2016}. Once the astrometry and photometry were measured in detector units, they were transformed into physical units accounting for the coronagraphic transmission, filter zero-points, and system distance \citep{Carter2023, Franson2024}, with systematic uncertainties on the star-behind-mask position (0.05\,pix) and stellar magnitude (0.02 mag) added in quadrature \citep{Girard2024}. The point source has a relative RA, Dec. $-0\farcs800\pm0\farcs004$, $-0\farcs777\pm0.006$ (i.e., separation $\alpha$ = $1\farcs115\pm0\farcs005$ and position angle PA = $225.84\pm0.26^\circ$, with a correlation $\rho=-0.5$). We derived an apparent magnitude of $19.68\pm0.07\,\mathrm{mag}$ in the F444W filter, and an upper limit of $19.0\,\mathrm{mag}$ in the F200W filter. The corresponding fluxes in SI units are $f_{\rm F444W}{=}3.92\times10^{-19}\pm2.5\times10^{-20}\,\mathrm{W/m^2/\mu m}$ and $f_{\rm F200W}{=}7.25\times10^{-19}\,\mathrm{W/m^2/\mu m}$, respectively. These fluxes correspond to absolute magnitudes of $M_\mathrm{F444W}{=}18.42\pm0.07$\,mag and $M_\mathrm{F200W}{\geq}17.74\,\mathrm{mag}$. The candidate point source was recovered in the F444W filter at a signal-to-noise ratio of $\mathrm{SNR}{=}5.7\,\sigma$, or a Bayesian evidence ratio of $\ln(B_{10}){=}\ln(Z_1/Z_0){=}115$ by comparing companion contrast to the annular contrast. 


\subsection{Planetary Prior Probability Compared Against Interloper Probability}

JWST is sensitive to background extragalactic sources far fainter than exoplanets in the Solar neighborhood. Previous high contrast imaging campaigns have ruled out sources surrounding 14~Her down to $15-30\,M_{\rm J}$ at 1\farcs0 \citep{Luhman2002, Patience2002, Carson2009, Rodigas2011, Durkan2016}. The red F200W-F444W color lower limit of the point source can be used to reject the vast majority of galactic contaminants. Since the probability of late T- or early Y-dwarf contaminants in such a compact field of view is vanishingly small \citep{2024ApJS..271...55K, Bogat2025}, we conclude that the dominant source of contamination in our images is extragalactic sources. 

In order to empirically estimate the number of contaminating sources that could confound a detection of 14 Her c, we calculated a space density of galaxies at all redshifts from JWST images that have similar observed characteristics as the planet. Following \citet{Bogat2025}, we acquired the JADES GOODS-S Deep High Level Science Product v2 from MAST\footnote{\doi{10.17909/8tdj-8n28}} \citep{Eisenstein2023, Rieke2023} covering 25 square arcminutes. 

We define a contaminant as a JADES source that appears point-like in the F444W filter, with a magnitude within the range of uncertainties of our candidate point source photometry, and with the F200W - F444W color lower limit of our point source. Specifically, we selected galaxies with AB magnitudes and colors comprised between $\mathrm{[F444W_{AB\,mag}]}=22.9\pm1.4$\,mag, and $\mathrm{[F200W_{AB\,mag} - F444W_{AB\,mag}]}=1.3$, and with a circular-equivalent size of FWHM$<4\,\mathrm{pix}$. We scaled the resulting number of sources by the sky area covered by the JADES GOODS-S Deep field to calculate the space density of galaxies with these characteristics. As our candidate source has an F444W magnitude well within the completeness limits of extragalactic surveys with JWST \footnote{JADES, and other JWST surveys, are 100\% complete to at least $\mathrm{[F444W_{AB\,mag}]}=28$\,mag, whereas our point source has $\mathrm{[F444W_{AB\,mag}]}=22.9\pm1.4$\,mag. See Figure 4 of \citet{2024A&A...691A.240M} for more details.}, and since the completeness corrected source counts for extragalactic surveys with JWST appear to agree well \citep[][their Figure 7]{2024A&A...691A.240M}, we do not expect any slight variation between JADES and other JWST deep fields to affect the robustness of the contaminant fraction estimate presented here.

For the predicted location of 14 Her c \citep{2023AJ....166...27B} and its 1, 2, and 3\,$\sigma$ contours (see leftmost panel of Figure \ref{fig:NIRCam_img}), we expect on average 0.003, 0.013, or 0.052 contaminating galaxies, respectively. Within 2\farcs0 of 14 Her A (see center and rightmost panel of Figure \ref{fig:NIRCam_img}), we expect on average 0.2 contaminating galaxies. Therefore, given the small area where the planet is expected, the existence of an extragalactic source with photometric properties that mimic those of the planet is extremely unlikely. A larger number of extended and blue JADES sources are expected within 2\farcs0, as evidenced by our detection of a resolved, blue galaxy to the SE of 14 Her A.

Balancing the extremely low likelihood that the point source is an extragalactic contaminant, and given that it is found in the predicted location, and with the expected luminosity for a giant planet of the age of the system (${\sim}18\,\mathrm{mag}$, see \S\ref{subsec:evol_and_atmo}), based on evolutionary models~\citep{Marley2021}, we conclude that the observed point source to the SW of the star is indeed 14 Her c.

This background contaminant analysis emphasizes the importance of prior information, especially system-specific dynamical priors, when attempting to confirm the existence of directly imaged companions in the infrared with a facility as sensitive as JWST \citep{Bogat2025}. The strong prior prediction for the location of the planet allows us to robustly claim a reliable detection despite the 0.2 probability of finding a ``14 Her c"-like source within the inner $2\farcs$0 of our image, and without requiring a follow-up observation to confirm orbital and/or common proper motion of the point source.



Limits at longer wavelengths (with MIRI or ALMA) are invaluable for reducing the probability a candidate point source is a background contaminant. For single epoch observations, including both NIRCam and MIRI coronagraphic imaging appears to be the best practice \citep[and, for nearby stars, may result in sensitivity to colder planets, see][]{Bowens-Rubin2025}. For example, In \citet{Lagrange2025}, the flux limit of a contaminating galaxy in the filter of interest (11.4\,\um) was 50\% across their $\sim10\farcs$ wide field of view. Only with deep upper limits from archival SPHERE (and especially) ALMA observations, were they able to reduce the contaminant probability in the 1\farcs5 immediately surrounding TWA~7 to 0.35\%.Similarly, a second epoch was required for the eventual confirmation of $\varepsilon$~Indi~A\,b \citep{Matthews2024}, where a low-significance early recovery of the candidate source in ground-based data was used to confirm the candidate's planetary nature. Strong dynamical priors, like those used to confirm TWA~7\,b \citep{Lagrange2025} or 14~Her~c, still contain a small probability of yielding false positives, but they allow for greater confidence in the initial characterization of expected planets. Future long wavelength imaging of 14 Her c with JWST would provide crucial color and temperature characterization, as well as incontrovertible confirmation of the planet's detection.

The spatial scales where JWST reaches its deepest contrasts are necessarily those where the risk of background contamination becomes large. Future studies, especially large coronagraphic surveys of young moving group stars with NIRCam dual-band coronagraphy (GO 4050, 5835, 6005, 7651) will need to rely heavily on ground-based upper limits from the literature, as well as any and all available dynamical limits to confirm cold planets that can currently only be detected with JWST.



\section{Analysis and Results}\label{sec:analysis_results}

\subsection{Orbit fitting}

\begin{figure}[!h]
    \centering
    \includegraphics[width=\linewidth]{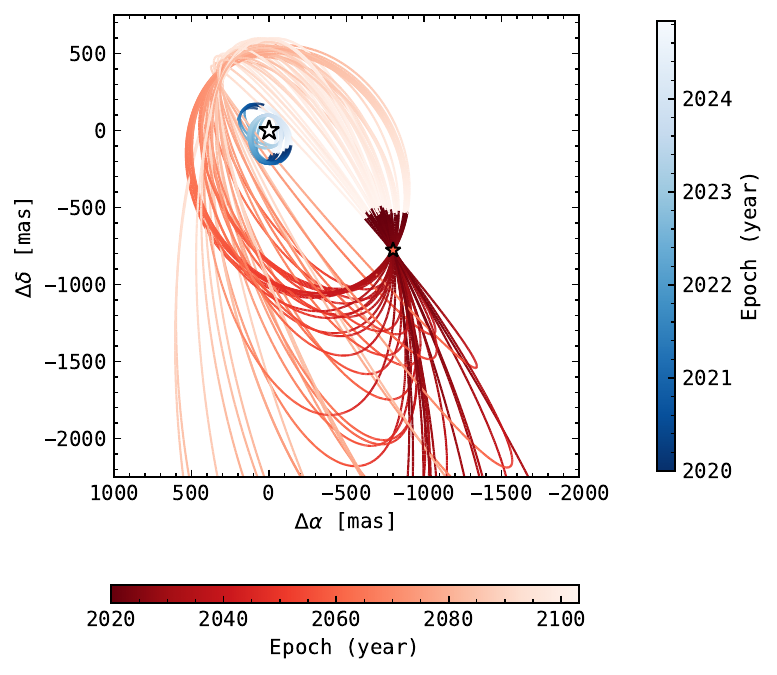}
    \caption{\textbf{Sky projection of the architecture of the 14 Her system}. The visual orbits for both planets, colored according to the time after the reference epoch 2020.0, $\mathrm{MJD}{=}58849.0$. The new NIRCam relative astrometry of 14~Her~c strongly constrains the relative orientation of the two planets' possible orbital planes.}
    \label{fig:orbs}
\end{figure}

We use our new JWST relative astrometry of 14 Her c to update the orbital parameters of the system, combining this data set with literature RVs from the ELODIE \citep{Naef2004}, HIRES \citep{Rosenthal2021}, HRS \citep{2023AJ....166...27B}, and APF \citep{Vogt2014} instruments spanning 25 years and absolute astrometry from Hipparcos and Gaia \citep{2018ApJS..239...31B}. We fit a total of 21 parameters with our 571 data points (3 from Gaia-Hipparcos absolute astrometry, 567 RV points, and 1 relative astrometry measurement from JWST) using two publicly-available orbit fitting codes: \texttt{orvara}\footnote{https://github.com/t-brandt/orvara} \citep{2021AJ....162..186B} and \texttt{orbitize!}\footnote{https://orbitize.readthedocs.io/en/latest/} \citep{2020AJ....159...89B}. Both of these orbit fitters use Bayesian inference with a Markov Chain Monte Carlo (MCMC) sampler to iteratively fit the data to Keplerian orbit models. Specifically, they use a Python implementation of a parallel-tempered version of the \texttt{emcee} \citep{2013PASP..125..306F} MCMC algorithm, namely \texttt{ptemcee} \citep{2021ascl.soft01006V}. These two codes differ in their implementation of implicit priors, their treatment of parallaxes, their flexibility incorporating arbitrary absolute astrometry, and their computational speed. 

Multi-planet systems are fit by iteratively solving two-body problems starting with the outermost companion. 
We used a Gaussian prior on the parallax of the star ($\pi{=}55.866\pm0.029$; \citep{GaiaCollaboration2022}), a log-flat prior on the primary mass and RV jitters (one per instrument), uninformative priors on projections of the eccentricity, mean longitude, and ascending node, and a $\sin i$ prior on orbital inclination to give equal probability to all possible orbital orientations. 

Our results with both orbit fitters use the same priors and data sets, and statistically agree on the accuracy and precision of the posterior probability distributions despite starting with different initial conditions and running for a different number of total steps, an indication that our results are robust and have converged. In Table~\ref{tab:orbitize_results}, we present the results from the longest run starting with agnostic initial conditions with \texttt{orbitize!}. Details on the initial conditions and code setup for each orbit fitter can be found in the Appendix~\ref{sec:orbitfit}.


Our measured relative astrometry of 14~Her~c establishes the tangential direction of the reflex motion that the planet exerts on its host star, helping to apportion the contribution of each planet to the total acceleration of 14~Her between the Hipparcos and Gaia measurements. The visual orbits for the two planets are shown in Figure \ref{fig:orbs}, and model fits to the stellar radial velocities are shown in Figure \ref{fig:rvs}. Table~\ref{tab:orbitize_results} records a summary of orbital parameter values. The new orbital solution for planet c gives a mass of $M_c = 7.9^{+1.6}_{-1.2}\,$\mjup, a semi-major axis of $a_c = 20.0^{+12.0}_{-4.9}\,\mathrm{au}$, and an eccentricity of $e_c = 0.52^{+0.16}_{-0.12}$, consistent with previous results \citep{2021ApJ...922L..43B, 2022ApJS..262...21F, 2023AJ....166...27B}. Two families of supplementary solutions for the inclination of 14 Her b arise from our orbit analysis: $i_b = 32.70^{+7.45}_{-4.01}\,^{\circ}$ and $i_b = 146.90^{+4.58}_{-9.74}\,^{\circ}$ driven by our lack of knowledge of the longitude of the ascending node for the inner planet ($\Omega_b$), which dictates the direction of a planet's motion along its orbit \citep[as noted by][for the case of 14~Her, either solution can be missed due to insufficient posterior sampling]{2024arXiv241214542F}. Inclusion of HST/FGS observations \citep{2023AJ....166...27B}, and follow-up JWST astrometry of 14 Her c and/or future direct imaging of 14 Her b with the Nancy Grace Roman Space Telescope ~\citep{2021SPIE11443E..3DG}, will provide crucial additional constraints on the relative orientations of the orbits, helping to further resolve the architecture of the system.

\subsection{Evolutionary and atmospheric modeling} \label{subsec:evol_and_atmo}

\begin{figure}[!h]
    \centering
    \includegraphics[width=\linewidth]{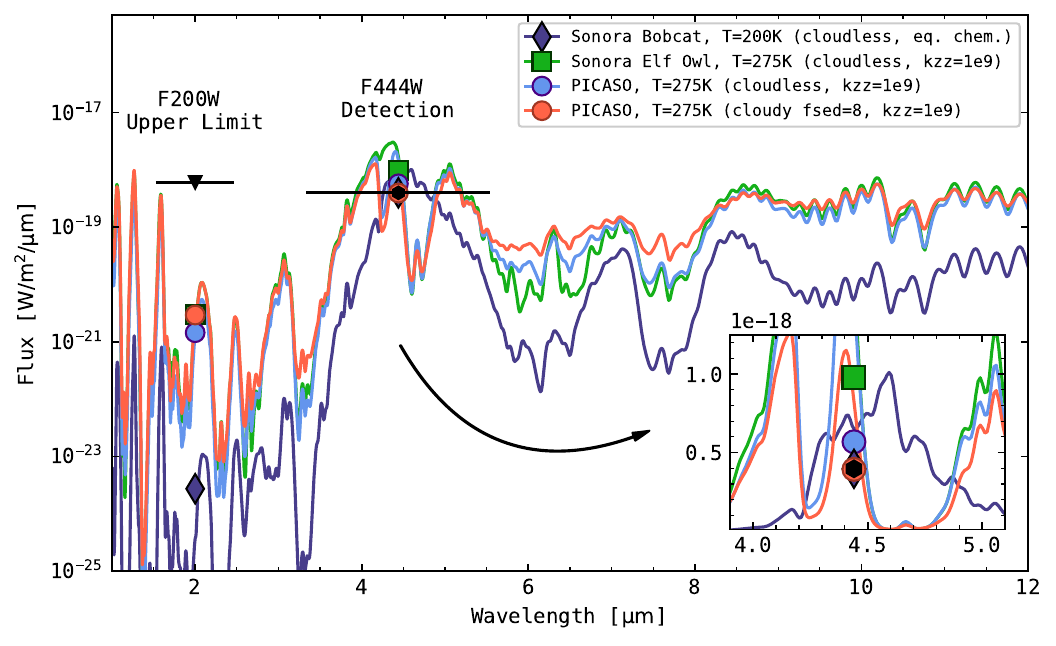}
    \caption{\textbf{Carbon disequilibrium chemistry and water clouds are likely in 14 Her c's atmosphere}. JWST/NIRCam photometry of 14~Her~c (F444W flux as a black hexagon, F200W upper limit as a black triangle) compared to atmospheric model spectra (colored lines) integrated over the filter bandpasses (colored circles, squares, diamonds). Publicly available Sonora Bobcat \citep{Marley2021} and Sonora Elf Owl \citep{2024ApJ...963...73M} models (in dark blue and green, respectively) are unable to reproduce the data with temperatures in agreement with evolutionary model predictions. Custom PICASO models with an updated treatment of carbon disequilibrium chemistry (light blue) brings the models in better alignment with the observations and evolutionary model predictions. Adding water ice clouds (red) to these custom atmospheric models brings the observations into agreement at $1\,\sigma$ with both the atmospheric and evolutionary models.}
    \label{fig:atm_models}
\end{figure}

Using our estimated dynamical mass for 14 Her c and the age posterior from the stellar characterization presented in \citep{2021ApJ...922L..43B}, we calculated fundamental parameters for the c planet with the \texttt{Sonora-Bobcat} cloudless substellar evolutionary models \citep{2021ApJ...920...85M}, assuming a metallicity of [Fe/H]=+0.5, representative of the stellar metallicity \citep{1999ApJ...511L.111G,2006AJ....131.3069L}. We linearly interpolated across the publicly available model grid and predict $T_{\rm eff}{=}300\pm30\,\mathrm{K}$, $\log(g){=}4.25\pm0.15$, $R_{\rm pl}{=}1.03\pm0.01\,\mathrm{R_{\rm j}}$, $\log(L/L_\odot){=}-7.1\pm0.2$. 
Integrating the NIRCam/F444W flux for a \texttt{Sonora-Bobcat} atmospheric model with these parameters, we predict an apparent magnitude of 17.98\,mag, which is 1.69\,mag brighter than our observed apparent magnitude in F444W (see Section~\ref{sec:obs}). Simultaneously fitting the \texttt{Sonora-Bobcat} atmospheric models to our F444W photometry and F200W limit, with $T_{\rm eff}$ as a free parameter and the other parameters fixed to their evolutionary values, we find $T_{\rm eff}{=}210\,\mathrm{K}$. This $\sim90\,$K tension can be reasonably attributed to carbon disequilibrium chemistry driven by vertical mixing, the presence of ice clouds, an enhanced metallicity \citep{2014ApJ...797...41Z,2023ApJ...950....8L} or some combination of these factors. Vertical mixing throughout the atmosphere can dredge up molecules that are formed at warm temperatures in the lower atmosphere and mix them into the cold, upper atmosphere faster than the chemical reaction timescale. This imbalance is most prominent among carbon and nitrogen molecules in objects colder than 500\,K~\citep{2014ApJ...797...41Z,2023ApJ...950....8L}. Carbon disequilibrium chemistry imparts deep CO and CO$_2$ absorption between $3-5\,\mu$m~\citep{2020AJ....160...63M,2023ApJ...951L..48B,2024ApJ...973..107B} in the spectra of substellar objects, even though the dominant carbon species at these temperatures should be CH$_4$ in chemical equilibrium~\citep{2023ApJ...950....8L,2014ApJ...797...41Z}. Clouds made of water ice particles can also absorb or scatter outbound light, and redistribute it to different wavelengths. To explore the effects of carbon disequilibrium chemistry and cloud formation on our measured F444W photometry, we generated a custom mini-grid of atmospheric models matching the metallicity of the star ([Fe/H] = +0.5), assuming a solar C/O, and including disequilibrium chemistry and water ice clouds based on the \texttt{Sonora Elf Owl} family of models \citep{2024ApJ...963...73M}. Details are described in Appendix~\ref{sec:atm_mod}.

Figure~\ref{fig:atm_models} shows that our JWST photometry in F444W is consistent with a cloudy atmosphere with carbon disequilibrium chemistry driven by strong vertical mixing \citep{2020AJ....160...63M,2023ApJ...951L..48B,2024ApJ...973..107B} at \teff = 275\,K and \logg = 4.25. This result puts 14~Her~c in the running for the coldest directly imaged exoplanet to date, a title currently held by Epsilon Indi Ab~\citep{2024Natur.633..789M}. Although these two planets have not yet both been imaged at equivalent wavelengths, the mid-infrared color of Epsilon Indi Ab similarly suggests a temperature of \teff$=275-300\,\mathrm{K}$.


\subsection{Dynamical simulations}\label{sec:supp_dynamics}

We explored the present-day dynamical evolution of the system by performing secular (i.e. orbit-averaged) three-body simulations using the publicly available \texttt{KozaiPy} software\footnote{https://github.com/djmunoz/kozaipy}. \texttt{KozaiPy} solves the equations of motion for three-body hierarchical systems ($a_b \ll a_c$) presented in \citet{2001ApJ...562.1012E, 2007ApJ...669.1298F} to octupole order $(a_b/a_c)^3$ \citep{naoz2013}. These packages include effects from equilibrium tides and general relativity, but as both planets are relatively far from the host star, such short-range effects are not impactful. In this limit, the semimajor axes of both planets remain fixed while mutual gravitational perturbations cause the eccentricities and inclinations to oscillate. We note that this level of approximation is strictly valid only for an inner test particle influenced by a massive, arbitrarily eccentric external perturber, and as such may not fully capture some relevant dynamics in the 14 Her system due to its similarly massive planets. There is no existing analytic prescription describing this regime, which is typically probed with direct \textit{N}-body simulations \citep{2013ApJ...779..166T}. We reserve detailed exploration of this for future work.

We sampled $\sim10,000$ parameter vectors from the posterior distribution derived from \texttt{orbitize!} and integrated them forward in time 10\,Myr, tracking the evolution of the orbital elements of both planets using orbit-averaged simulations assuming a hierarchical approximation ($a_b \ll a_c$). All orbital solutions show substantial eccentricity and inclination oscillations for both planets as a result of strong secular perturbations. Figure \ref{fig:dynamical simulations} shows the distribution of the eccentricity oscillation amplitude of 14 Her b $\Delta e_b \equiv \max{e_b} - \min{e_b}$, as well as examples of two representative simulations -- one characterized by moderate eccentricity and inclination oscillations, and one by much stronger oscillations.

\section{Discussion}\label{sec:discussion}

\subsection{On the mutual inclination of the 14 Her planets}


With our new orbital solutions, we revise the mutual inclination angle $\Theta$ between the two orbital planes, defined as $$\cos{\Theta}=\cos{i_b}\cos{i_c} + \sin{i_b}\sin{i_c}\cos{(\Omega_b- \Omega_c)}$$ in e.g.,~\citealt{2021ApJ...922L..43B,2019ApJ...883...22C}. This results in a bimodal posterior distribution of mutual inclinations of $\Theta = 32^{+13.6}_{-15.1}\,^{\circ}$ and $\Theta = 145.0^{{+15.8}\,\circ}_{-11.1}$, for $i_b$ greater than or less than $90^\circ$, respectively. These values represent roughly a factor of two improvement in precision compared to the previous estimate of the mutual inclination before our relative astrometry measurement of 14 Her c with JWST \citep[$\Theta = 96.3^{+36.8}_{-29.1}\,^{\circ}$,][]{2021ApJ...922L..43B}. The nonzero mutual inclination between the orbits of 14 Her b and c and large eccentricities are in stark contrast with the planets in our own Solar System, as well as younger, directly imaged multiplanet systems like HR~8799 \citep{2008Sci...322.1348M}, $\beta$~Pictoris \citep{2010Sci...329...57L}, and HD~206893 \citep{Hinkley2023} that have multiple planets on (roughly) coplanar, low eccentricity orbits~\citep{2020AJ....159...63B}. 

\begin{figure}
    \centering
    \includegraphics[width=\linewidth]{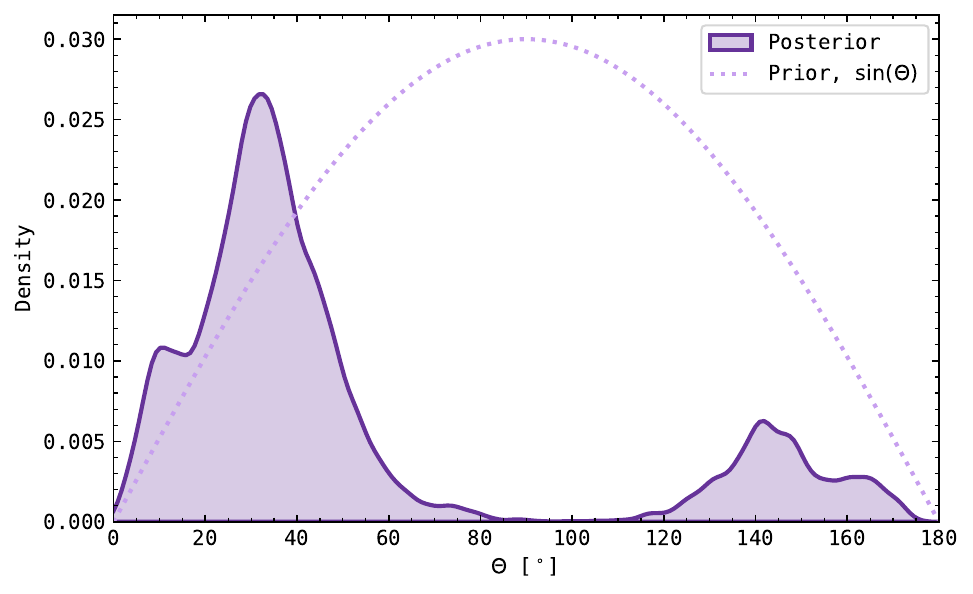}
    \caption{
    \textbf{The posterior distribution of mutual inclination angles $\Theta$ between the orbits of 14~Her~b and c}, compared to the Sine prior. The constraint on the mutual inclination is driven by the likelihood function and not the prior.
    }
    \label{fig:theta}
\end{figure}

\subsection{Potential for disequilibrium chemistry in the atmosphere of 14~Her~c}

14 Her c is over a million times fainter than its host star at $4.4\,\mu$m (a contrast of $9.6\pm0.5\times10^{-7}$, see Figure \ref{fig:cc}), and roughly one magnitude fainter than the coldest isolated brown dwarf known, WISE J0855$-$0714~
\citep{2024ApJ...977L..49R}~
in the comparable Spitzer filter centered at $4.5\,\mu$m. The 90\,K degree tension between the temperatures from evolutionary and spectral models is $>3\,\sigma$ significant, suggesting the presence of
additional sources of $4.4\,\mu$m opacity in the planet's atmosphere such as clouds or disequilibrium chemistry.



Our custom model fits imply that 14 Her c should have a deep CO$_2$ absorption feature at $4.2\,\mu$m \citep{Beiler2024}, which can be tested using follow-up JWST coronagraphic images with narrower bandwidth filters \citep{Balmer2025b}. Similarly, our photometry implies that other cold planets like 14 Her c are likely slightly fainter at $4.4\,\mu$m than expected for their mass and age. Future JWST observations of 14 Her c with multi-band photometry across the near and mid-infrared could feasibly quantify the thermal emission, cloud opacity slope, and CO$_2$, CO, and NH$_3$ absorption feature strengths predicted by these models. Direct spectroscopy with JWST may be able to further refine the molecular abundances in the planet's atmosphere, although this will be challenging due to the planet's high contrast. These observations would provide the first glimpses into the complexity of this field-age, giant exoplanet atmosphere as a benchmark for other cold exoplanets and Y-type brown dwarfs, yielding new insights into the physics and chemistry occurring in low temperature giant planet atmospheres.

\subsection{A cold planet in a dynamically hot system}

The unusual present configuration of this system is not representative of most observed multi-planet systems and instead points to past and ongoing dynamical interactions. The results from our dynamical simulations (see Section~\ref{sec:supp_dynamics}) indicate that both planets in the 14 Her system are currently experiencing significant secular eccentricity and inclination oscillations. While it is impossible to directly compare the magnitude of these excursions to the exoplanet population at large (as such comprehensive detailed dynamical modeling does not exist for all systems), an illustrative comparison can be made with the Solar System. The largest amplitude eccentricity excursions experienced by a Solar System planet are of order $\sim0.1$ by Mercury, whereas 14 Her b, on average, experiences oscillations with nearly five times this amplitude. This behavior makes 14 Her the only currently known dynamically ``hot" multi-planet system with an imaged planet\footnote{For example, the dynamically ``cold" directly imaged multi-planet system $\beta$ Pictoris bc likely experiences secular eccentricity oscillations of order $\Delta e<0.2$ \citep{Lacour2021}, and the HR 8799 bcde system likely experiences oscillations of order $\Delta e <0.05$ \citep{Wang2018}.}. However, even though the orbital parameters vary periodically, the system is stable on this timescale.   


Planet-planet scattering, including a planetary ejection, was proposed as a potential hypothesis to explain the origin of this peculiar system architecture~\citep{2021ApJ...922L..43B}. With our new orbital constraints, future N-body simulations exploring this and other initial scenarios will be key to gauge the feasibility of this mechanism in shaping the configuration of this system and its relative occurrence among the long-period exoplanet population. Ejections of giant planets may then be a common avenue to produce 
the observed population of ``rogue'' planets in our galaxy \citep{Sumi2023}. 

\begin{figure}
    \centering
    \includegraphics[width=\linewidth]{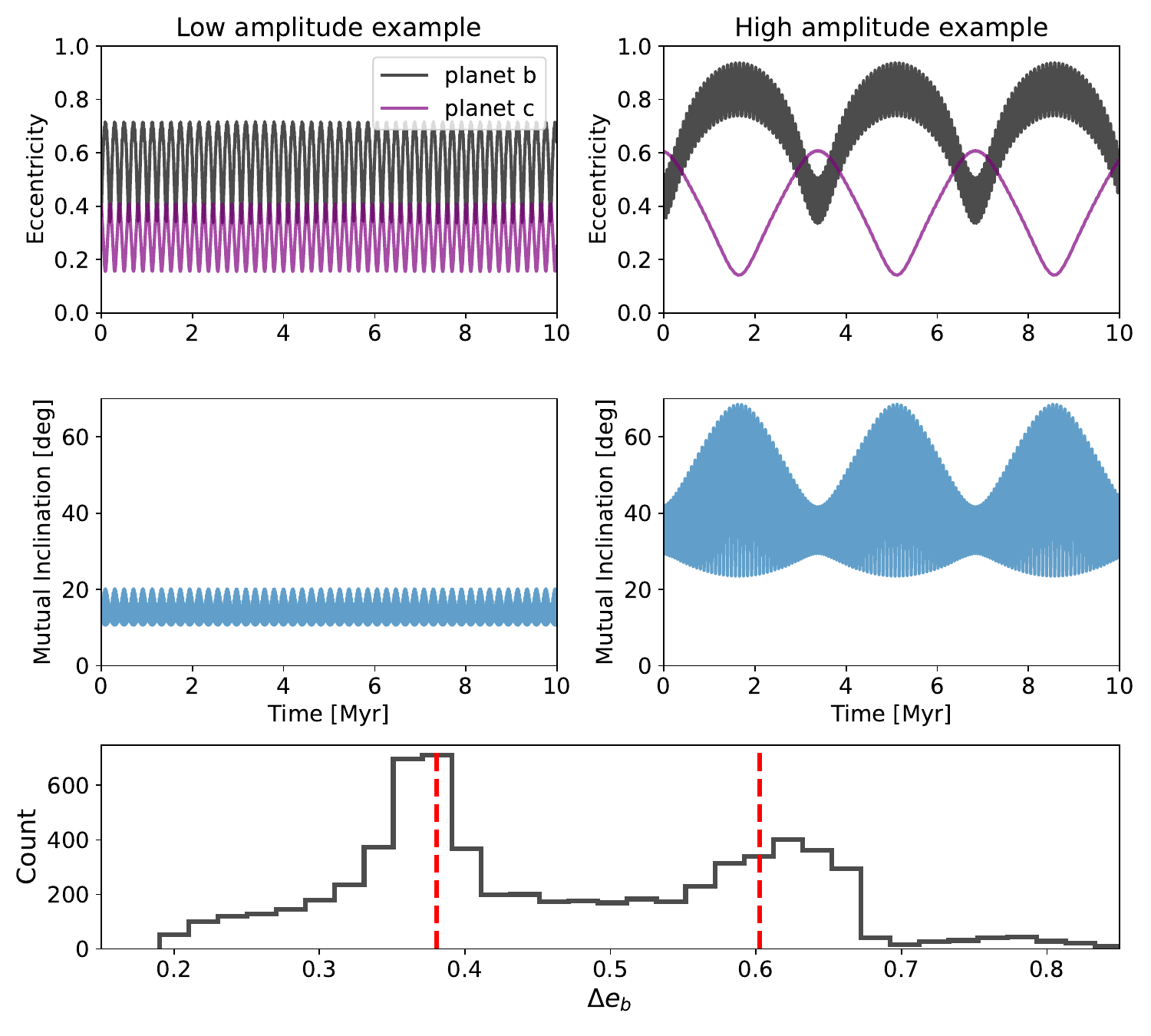}
    \caption{\textbf{Example dynamical simulations}, initialized from different parameter vectors from the \texttt{orbitize!} posterior distribution. The left and right columns show examples of lower and higher amplitudes of secular oscillations, as seen in the eccentricity variations of both planets and the mutual inclination variations. The histogram in the bottom panel shows the distribution of $\Delta e_b$ across all simulations, and the values corresponding to the two example simulations are shown with the vertical red lines.}
    \label{fig:dynamical simulations}
\end{figure}

\section{Conclusions}\label{sec:conclusions}

\par In this letter we present the first direct images of 14 Her c. The key takeaways of our study are as follows:
\begin{enumerate}
\item Coronagraphic imaging of 14 Her with JWST/NIRCam achieved deep contrasts ($<10^{-6}$) at thermal wavelengths (4.5\um) and spatial scales between 15-150\,au. These observations detected the known planet 14 Her c at $>5\,\sigma$ confidence.
\item The planet appears within $1\,\sigma$ of its predicted location from orbits fit to previous absolute astrometry and radial velocity measurements. The probability of a chance alignment with a background source are negligible compared to the likelihood the source is 14 Her c. 
\item 14~Her~c (\teff${\sim}275{-}300\,\mathrm{K}$) rivals $\epsilon$~Indi~Ab \citep[${\sim}275{-}300\,\mathrm{K}$,][]{Matthews2024} and TWA~7~b \citep[${\sim}315\,\mathrm{K}$,][]{Lagrange2025} as one of the coldest planets yet directly imaged, and with $\epsilon$~Indi~Ab \citep[$\sim3.5\,$Gyr,][]{Matthews2024} as one of the oldest ($\sim4.6\,$Gyr). 
\item Temperature estimates for 14 Her c disagree slightly within 90\,K or $1\,\sigma$ between the estimate from dynamical mass and age and that from the F444W photometry. This tension between estimates could be explained by the presence of carbon disequilibrium chemistry and/or water ice clouds in the atmosphere. See Figure~\ref{fig:atm_models} for atmospheric modeling. 
\item The misalignment between orbits identified in~\citet{2021ApJ...922L..43B} remains. With our new detection, we have refined the parameters of both planetary orbits in the system and constrained the misalignment more precisely. Only five other multi-planet systems with measured orbits exhibit a significant misalignment, and so far 14 Her is the only one with a directly imaged planet. 
\item The unusual system architecture of 14 Her c is suggestive of dramatic past and present dynamical interactions. Our preliminary dynamical simulations indicate ongoing secular eccentricity and inclination oscillations with large eccentricity excursions roughly 5 times those of Mercury.
\end{enumerate}

This system is a unique benchmark to describe the dynamical evolution of multi-planet systems and the atmospheric properties of Jupiter analogs. Future multi-band imaging or cross correlation spectroscopy with JWST can provide a glimpse into the spectral energy distribution and the atmospheric composition of this planet. 14 Her is an exciting target for the Coronagraphic Instrument aboard the Roman Space Telescope. Precisely constraining the location of the inner planet will establish both planetary orbits in the system, the degree of orbital misalignment, and exclude a subpopulation of orbital solutions to improve our knowledge on the ongoing dynamical interactions of the system. 


\begin{acknowledgments}

We thank an anonymous reviewer for their review, which improved the quality of this manuscript.
D.B.G., W.O.B, and M.G. acknowledge support from Space Telescope Science Institute JWST grant No. JWST-GO-03337.
J.M. is supported by the NSF Graduate Research Fellowship.
This work was performed in part using high-performance computing equipment at Amherst College obtained under National Science Foundation (NSF) Grant No. 2117377. Part of this work was carried out at the Advanced Research Computing at Hopkins (ARCH) core facility (rockfish.jhu.edu), which is supported by the National Science Foundation (NSF) Grant No. OAC1920103.

W.O.B. thanks E. Bogat, J. I. Adams Redai, R. Bendahan-West, R. Ferrer-Chavez, A. James, and R. Kane for helpful discussions.

\end{acknowledgments}

\begin{contribution}

D.B.G. and W.O.B. planned and proposed the James Webb Space Telescope observations, led the collaboration, and wrote the manuscript.
W.O.B. and L.P. led the JWST image processing, starlight subtraction, photometric and astrometric measurements. 
D.B.G., W.O.B., M.G., and T.D.B. led the orbit fitting with both \texttt{orvara} and \texttt{orbitize!}. 
D.B.G., W.O.B. estimated fundamental parameters with evolutionary models. 
S.M., T.L., M.R., B.B. designed and ran secular dynamical simulations.
J.M., C.M., B.L. developed a mini-grid of atmospheric models and fit the measured JWST photometry to them. 
J.G. advised on observation planning for the NIRCam coronagraph.
E.C.M., A.C., B.P.B., J.F., C.F., E.R. provided comments and expertise throughout the analysis.


\end{contribution}

\facilities{JWST(NIRCam), OHP:1.93m(ELODIE), Keck:I(HIRES), Smith(Cross-Dispersed Echelle Spectrograph), APF, HIPPARCOS, Gaia}

\software{astropy, pyKLIP, spaceKLIP, orbitize!, orvara, KozaiPY, PICASO}

\appendix

\section{Image reduction}\label{sec:img_red}

The uncalibrated Stage 0 (*uncal.fits) data products were acquired from the Barbara A. Mikulski Archive for Space Telescopes (MAST) \footnote{
The data described here may be obtained from the MAST archive at\dataset[doi:10.17909/gx61-s247]{https://dx.doi.org/10.17909/gx61-s247}} and processed with \texttt{spaceKLIP}\footnote{https://spaceklip.readthedocs.io/en/latest/}, a community developed pipeline for high contrast imaging with JWST \citep{Kammerer2022}. \texttt{spaceKLIP} wraps the \texttt{jwst} pipeline \citep{Bushouse2023} for basic data processing steps with modifications for coronagraphic imaging reduction. Data reduction via \texttt{spaceKLIP} for observations using the F444W filter and \texttt{MASK335R} coronagraph has been described at length \citep{Kammerer2022, Carter2023, Franson2024, Kammerer2024, Balmer2025b}. %
We used \texttt{spaceKLIP} v2.1, via github commit \#11df3a1, and \texttt{jwst} v1.18; the calibration files were from CRDS v12.1.5 and the \texttt{jwst\_1364.pmap} CRDS context\footnote{https://jwst-crds.stsci.edu/display\_context\_history/}.
Briefly, the data were fit ``up the ramp" using the ``Likely" algorithm described in \citep{Brandt2024a,Brandt2024b} and transformed from Stage 0 images into Stage 2 (*calints.fits) images, using a jump threshold of 4 and 4 pseudo-reference pixels on all sides of the subarrays; this process collates the frames (non-destructive reads and averages of the detector) into integrations, which end with destructive readouts of the detector, while using the differences between groups to correct for cosmic rays and other artifacts. The 1/f noise typical of thermal imaging with HgCdTe arrays was mitigated using a median filter computed along each column. We skipped dark current subtraction following \citep{Carter2023}. 
Pixels flagged by the \texttt{jwst} pipeline as well as $5\,\sigma$ outliers detected post facto using sigma clipping were replaced with a 2-dimensional interpolation based on a 9-pixel kernel. We also identified additional pixels affected by cosmic rays by flagging pixels with significant temporal flux variations across integrations and replaced them by their temporal median. Subsequently, the images were blurred above the Nyquist sampling criterion, using a Gaussian with a FWHM of 2.70 for the F444W/LW detector and 2.77 for the F200W/SW detector. These values ensure that sharp features in the data do not create artifacts when Fourier shifts are applied to the undersampled images. The position of the star behind the coronagraph was estimated by fitting a model coronagraphic PSF from \texttt{webbpsf\_ext}\footnote{https://github.com/JarronL/webbpsf\_ext} to the first science integration. All subsequent images were shifted by this initial offset. Images were cross-correlated to the first science integration, and these small shifts were applied to each integration to center the entire observing sequence to the first science integration. The recovered shifts were equivalent to and reproduced the small grid dither sequence that was commanded.

\subsection{Starlight subtraction and contrast estimation}

\par The residual starlight in the coronagraphic images was modeled and removed using the Karhunen--Lo\`{e}ve Image Projection (KLIP) algorithm \citep{Soummer2012} through the Python implementation \texttt{pyKLIP}\footnote{https://pyklip.readthedocs.io/en/latest/} \citep{Wang2015} wrapped by \texttt{spaceKLIP}. This algorithm uses Principal Component Analysis (PCA) to create an orthogonal basis set of eigenimages from the reference images. The science images are projected onto this basis to isolate and remove the contribution from the stellar PSF common between the reference and science images. This can be done on annular or radial subsections of the data by specifying a number of \texttt{annuli} and \texttt{subsections} in which to split the data, while the number of eigenimages used to construct the PSF model are specified with the \texttt{numbasis} parameter (sometimes referred to as the number of ``KL modes").  

\par We used ADI and RDI to create our model PSF and subtracted the starlight with $\texttt{annulus}{=}1$, $\texttt{subsections}{=}1$, and $\texttt{numbasis}{\in}[1,2,3,4,5,6,7,8,9,10, 25, 50, 100]$. In the F200W filter, an extended source (a background galaxy) is visible in all KL modes at a separation of 1\farcs5 to the SE, and becomes very apparent by the $6^{th}$ KL mode. In the F444W filter, this same galaxy appears by the $5^{th}$ KL mode subtraction, and another source, a point source, becomes apparent to the SW. By the $10^{th}$ KL mode, most other speckles in the image have been suppressed, leaving the galaxy at 1\farcs5 to the SE in both filters, a few faint, extended sources beyond 2\farcs0 (one at 4\farcs0 to the NE, and one at 8\farcs9 to the N), and the point source to the SW in F444W. We detected no other point sources in either filter. We verified that the point source appears in both roll angles and rotates with the astrophysical scene, not the detector frame. The signal-to-noise ratio of the source climbs from ${<}1$ to $5.7$ between \texttt{numbasis} 1 and 5, and plateaus about this value for \texttt{numbasis}${>}5$. We tested the robustness of the source to the starlight subtraction by varying \texttt{KLIP} parameters, and found that the point source persists for $\texttt{annuli}{\in}[2{-}16]$ and $\texttt{subsections}{\in}[2{-}8]$, \texttt{numbasis}${>}5$. The significance of the source does not vary dramatically with the choice of KLIP parameter, but is greatest for the simplest, least ``aggressive" choice in parameters, $\texttt{annuli}{=}1$, $\texttt{subsections}{=}1$ because there is the least self-subtraction for these parameters \citep{AdamsRedai2023}. We also verified that the source persists, albeit at lower SNR, in subsets of the science integrations; we were able to detect the source at ${\gtrsim}4\,\sigma$ in up to a third of the total dataset, regardless of which frames were selected. Since the contrast and position of the point source agrees well with the mass and position predicted for the epoch of observation based on orbits for 14 Her c from \citep{2021ApJ...922L..43B} and \citep{2023AJ....166...27B}, we proceed to estimate its astrometry and photometry using the $\texttt{annuli}{=}1$, $\texttt{subsections}{=}1$, $\texttt{numbasis}{=}25$ PSF subtracted images (Figure \ref{fig:NIRCam_img}).

\begin{figure}
    \centering
    \includegraphics[width=\linewidth]{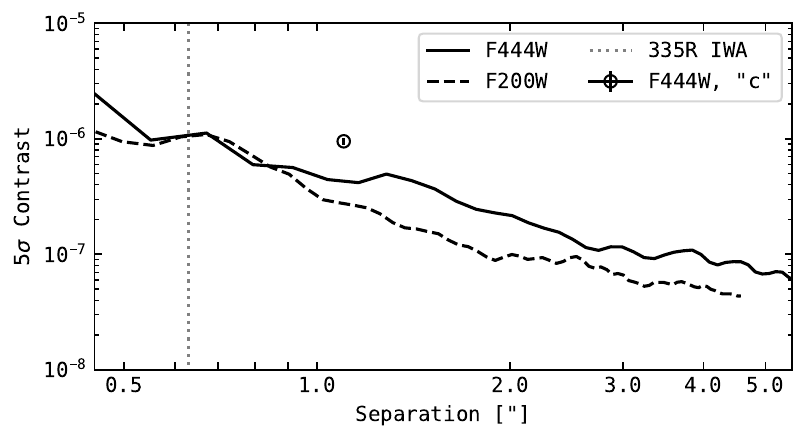}
    \caption{\textbf{Contrast curve for the NIRCam observations of 14~Her using the \texttt{335R} coronagraph.} The F444W and F200W $5\,\sigma$ contrast curves are shown as solid and dashed lines respectively. The likely detection of 14~Her~c at ${\sim}1\farcs1$ is plotted as an black circle and error bar. The nominal \texttt{335R} inner working angle (``IWA"), where the coronagraph transmission reaches 50\%, is indicated by a vertical dashed line. \textit{Takeaway}: Our detection of 14~Her~c is statistically significant.}
    \label{fig:cc}
\end{figure}

\subsection{Forward model subtraction of both sources}

\begin{figure}
    \centering

    \gridline{
    \fig{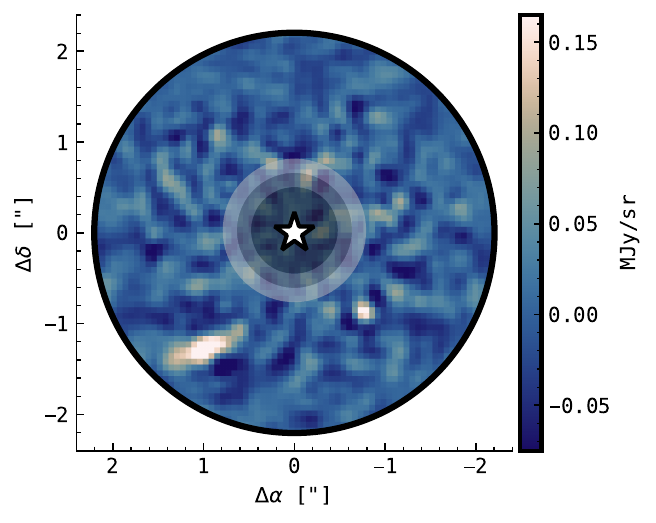}{0.31\textwidth}{(a)}
    \fig{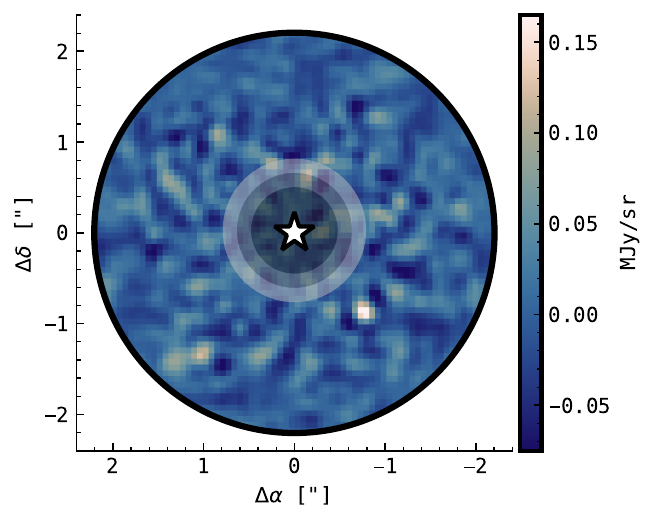}{0.31\textwidth}{(b)}
    \fig{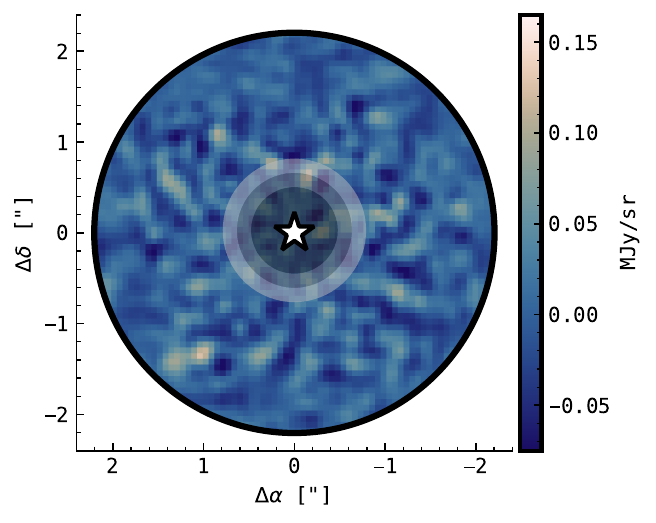}{0.31\textwidth}{(c)}
    }
    
    \caption{\textbf{Forward modeling of sources in the NIRCam F444W image.} (a) Reproduction of the PSF subtracted image in Figure \ref{fig:NIRCam_img}. The two prominent sources are an extended source (a galaxy) to the SE and a point source (14~Her~c) to the SW. (b) The residuals after the subtraction of a simple forward model of the background galaxy, a \texttt{webbpsf} point source convolved by a 2D Gaussian. Small (10\%) residuals to the SE of the center of the galaxy are apparent. The visual significance of 14~Her~c is apparent, as the galaxy has been suppressed. (c) The residuals of the image after the point source forward model for 14~Her~c has been subtracted.  }
    \label{fig:NIRCam_img_sourcesub}
\end{figure}

The forward modeling of the point source, 14~Her~c, is described in \S\ref{sec:obs}.1. In this section, we briefly describe a simplistic forward model for the extended source, the background galaxy, in the F444W filter and present the residual images for the forward modeling. The procedure for modeling the extended source follows from the point source forward model, except that at each step the \texttt{webbpsf} model is convolved with a 2D Gaussian parameterized by a FWHM in the x and y directions and a correlation term $\theta$. This model does not fully capture the structure of the source, as residuals are apparent to the SE of the central source, but serves to describe it at first order. Figure \ref{fig:NIRCam_img_sourcesub} illustrates the F444W image (Figure \ref{fig:NIRCam_img}), the same image with the background galaxy subtracted, and with both sources subtracted. 

\section{Orbit fitting}\label{sec:orbitfit}

\subsection{\texttt{orvara}}

For an initial \texttt{orvara} run, we used all three datasets with 100 walkers over 30 temperatures, 1 million steps per walker (i.e., 500,000 steps for each planet) and saved every $50^{th}$ step for inference, starting with the median parameter values from the posteriors published in \citep{2021ApJ...922L..43B} as initial conditions. We constrained the RV jitter for each instrument to a $10^{-5}-10$\,m/s range. The total chain length per walker was 20,000 out of which the first 500 steps were discarded as burn-in. 

The single epoch of JWST relative astrometry for the 14 Her c planet helped to strongly constrain its orbital parameters. However, our lack of constraint on the longitude of the ascending node for 14 Her b produced a bimodal distribution for its orbital inclination. About 2/3 of the orbit posteriors were consistent with an inclination of b lower than $90^{\circ}$, whereas the remaining 1/3 of orbits had inclinations for b greater than $90^{\circ}$. To explore this degeneracy, we ran new orbit fits to produce longer chains of 3 million steps per walker divided evenly over the two planets, using 100 walkers over 30 temperatures. On separate runs, we tried each family of inclinations for b and the associated parameters for b and c as initial conditions. The total chain length used for inference on each run had 60,000 steps with 3000 steps discarded as burn-in, both with a mean acceptance fraction of 0.029. 

It appears as if the initial run with 1 million steps had not yet converged and based on the starting conditions, which initialized at an inclination of $32.7^{+3.2}_{-5.3}\,^{\circ}$ for b \citep{2021ApJ...922L..43B}, gave preference to this family of solutions while only starting to explore the family of supplementary inclinations greater than $90^{\circ}$. The subsequent longer runs reversed the proportion of orbits of 14 Her b with each inclination, with only 1/5 of posteriors consistent with a $i_b<90^{\circ}$ regardless of the initial conditions of b inclinations (i.e., below or above $90^{\circ}$).

\subsection{\texttt{orbitize!}}
We corroborated our results with an independent orbit fit using \texttt{orbitize!}. We set normally distributed priors on the host star mass ($M_{*}{=}0.98\pm0.04$; \citep{2021ApJ...922L..43B}) and the system parallax ($\pi{=}55.866\pm0.029$; \citep{GaiaCollaboration2022}). We adopted uniform priors between $-5$ and $5\,\mathrm{km/s}$ for the RV zero points, and log uniform priors between $10^{-5}$ and $10^{2}\,\mathrm{km/s}$ for the RV jitter nuisance parameters. We restrict the semi-major axes for the inner and outer planets not to cross, placing log-uniform priors between 0.1 and 5\,au for the inner, and between 5 and 500\,au for the outer. Similarly, we place log-uniform priors on the planet masses ranging from 0.2 to 20 Jupiter masses. We adopted uninformative priors on the eccentricity and the angles describing the visual orbital elements for the two planets (Table 1; \citep{2020AJ....159...89B}).

\begin{table}[]
    \centering
    \footnotesize
    \begin{tabular}{cccccc}
    Parameter & Description & Units & Prior & Max. A-posteriori & Median and 68\% CI \\
    \hline \hline 

$a_{\rm b}$	&  Semi-major axis  &  au &  Log-uniform [0.1, 5] & 2.839   & $2.839^{+0.039}_{-0.041}$ \\
$e_{\rm b}$	&  Eccentricity  &  ... & Uniform [0,1] & 0.3683  & $0.3683^{+0.0029}_{-0.0029}$ \\
$i_{\rm b}$	&  Inclination  &  degree & Sine [0,180] & 147.8 & $146.7^{+4.584}_{-9.740}$ \\  
$i_{\rm b}$ $^\star$	&  Inclination  &  degree & Sine [0,180] & 29.79 & $32.66^{+7.448}_{-4.011}$ \\  
$\Omega_{\rm b}$	&  Longitude of the Ascending Node  &  degree & Uniform [0,360] & 202.7 & $197.8^{+4.240}_{-5.615}$ \\  
$\Omega_{\rm b}$ $^\star$ &  Longitude of the Ascending Node  &  degree & Uniform [0,360] & 44.12 & $44.69^{+3.438}_{-4.011}$ \\  
$\tau_{b}$	&  Relative Periastron $^\ddagger$  &  ... & Uniform [0,1] & 0.7670   & $0.7672^{+0.0013}_{-0.0013}$ \\   
$a_{\rm c}$	&  Semi-major axis  &  au &  Log-uniform [5, 500] & 15.1 & $20.0^{+12.0}_{-4.9}$ \\ 
$e_{\rm c}$	&  Eccentricity  &  ... & Uniform [0,1] & 0.40  & $0.52^{+0.16}_{-0.12}$ \\ 
$i_{\rm c}$	&  Inclination  &  degree & Sine [0,180] & 114.0   & $116.3^{+24.64}_{-9.167}$ \\
$\omega_{\rm c}$	&  Argument of periastron $^\dagger$  &  degree & Uniform [0,360] & 173.0    & $172.5^{+4.011}_{-4.584}$ \\   
$\Omega_{\rm c}$	&  Longitude of the Ascending Node  &  degree & Uniform [0,360] & 206.8    & $205.1^{+7.448}_{-10.31}$ \\
$\tau_{c}$	&  Relative Periastron $^\ddagger$  &  ... & Uniform [0,1] & 0.66  & $0.78^{+0.11}_{-0.12}$ \\   
$\pi$   &  Parallax	    &  mas    & $\mathcal{N}(55.866,0.029)$ & 55.871   & $55.866^{+0.028}_{-0.029}$ \\ 
$\gamma_{\rm APF}$	    & RV ZP &  km/s &  Uniform & 0.044  & $0.031^{+0.013}_{-0.017}$ \\  
$\sigma_{\rm APF}$	    & RV jitter &  km/s &  Log-uniform & 0.00345 & $0.00347^{+0.00027}_{-0.00023}$ \\   
$\gamma_{\rm ELODIE}$	& RV ZP &  km/s & Uniform & -0.028 & $-0.041^{+0.013}_{-0.017}$ \\ 
$\sigma_{\rm ELODIE}$	& RV jitter &  km/s & Log-uniform & 0.007 & $0.007^{+0.001}_{-0.001}$ \\   
$\gamma_{\rm HIRES-POST}$  & RV ZP &  km/s & Uniform & -0.0041 & $-0.017^{+0.013}_{-0.016}$ \\
$\sigma_{\rm HIRES-POST}$  & RV jitter &  km/s & Log-uniform & 0.0031 & $0.0031^{+0.0002}_{-0.0002}$ \\
$\gamma_{\rm HIRES-PRE}$	& RV ZP &  km/s & Uniform & 0.0032 & $-0.010^{+0.013}_{-0.017}$ \\    
$\sigma_{\rm HIRES-PRE}$	& RV jitter &  km/s & Log-uniform & 0.0027 & $0.0028^{+0.0004 }_{-0.0004}$ \\   
$\gamma_{\rm HRS}$	 & RV ZP &  km/s  & Uniform & -0.032 & $-0.045^{+0.013}_{-0.017}$ \\
$\sigma_{\rm HRS}$	 & RV jitter &  km/s  & Log-uniform & 3.9e-05 & $0.00015^{+0.00084}_{-0.00013}$ \\
$\mathcal{M}_{\rm b}$     & Mass of b   & $M_{\rm J}$  & Log-uniform [0.2,20] & 9.2 & $8.9^{+1.3}_{-1.7}$ \\  
$\mathcal{M}_{\rm c}$     & Mass of c   & $M_{\rm \odot}$  & Log-uniform [0.2,20] & 7.5 & $7.9^{+1.6}_{-1.2}$ \\ 
$\mathcal{M}_{\rm A}$     & Mass of A   & $M_{\rm \odot}$  & $\mathcal{N}(0.98,0.04)$ & 0.965  & $0.97^{+0.04}_{-0.04}$ \\
\hline
$\Theta$ & Mutual inclination & degree & ... & 31.51 & $32.09^{+13.75}_{-14.90}$ \\
$\Theta$ $^*$ & Mutual inclination & degree & ... & 141.5 & $145.0^{+16.04}_{-10.89}$

    \end{tabular}
    \caption{\textbf{Orbit fit results}. The results of the longest MCMC run using \texttt{orbitize!} was adopted here. $^*$ $i_b<90^\circ$ mode solutions for the orbit of b, with roughly 3 to 1 probability compared to the  mode containing $i_b<90^\circ$ solutions. $^\dagger$ The argument of periastron of the planet's orbit, not the star's orbit, about the center of mass. $^\ddagger$ Defined as $\tau=\frac{t_p-t_{\rm ref}}{P}$, where $t_p$ is the time of periastron passage and $t_{\rm ref}=58849.0$ MJD.}
    \label{tab:orbitize_results}
\end{table}

\begin{figure}
    \centering
    \includegraphics[width=0.8\linewidth]{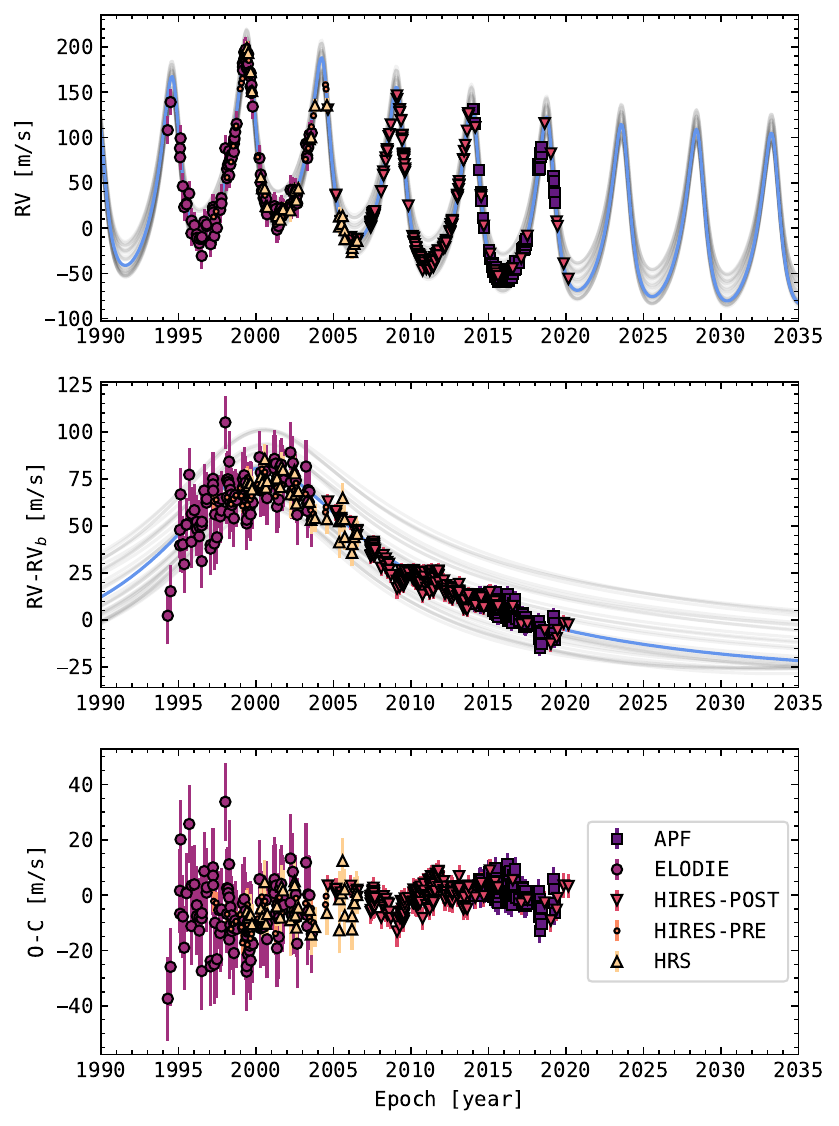}
    \caption{\textbf{The stellar radial velocities driven by the orbits of both planets}, as measured by various instruments. The $\sim5\,\mathrm{yr}$ modulation from 14~Her~b is apparent, as is the long term variation due to 14~Her~c, although there is not complete coverage of the outer planet's orbit.}
    \label{fig:rvs}
\end{figure}

\begin{figure}
    \centering
    \includegraphics[width=\linewidth]{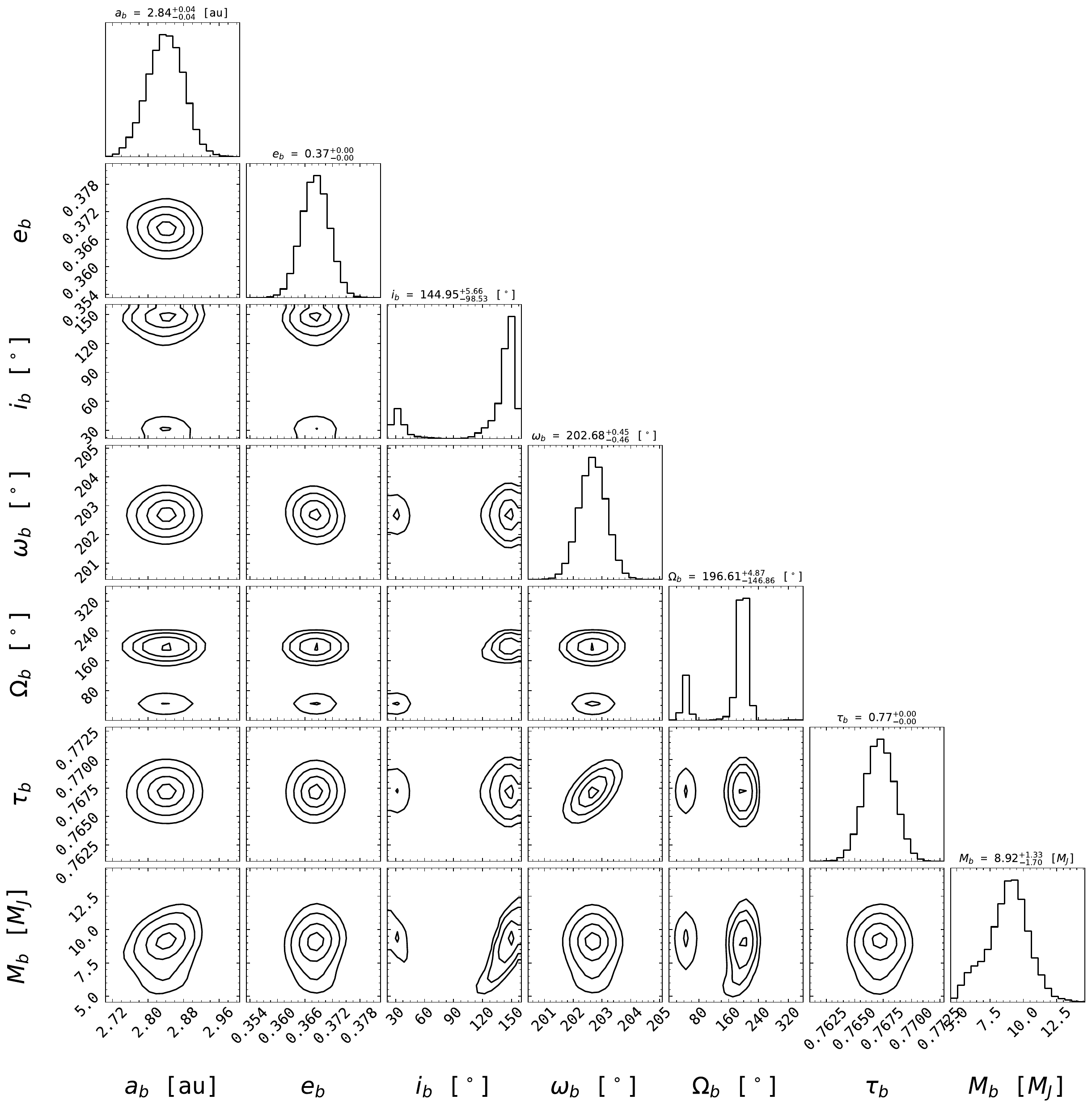}
    \caption{\textbf{Posterior distribution on the orbital elements of 14~Her~b}, from the \texttt{orbitize!} orbit fit. Two solutions for the planet's inclination $i_b$ and longitude of the ascending node $\Omega_b$ arise due to the ambiguity of the direction of the planet's orbit.}
    \label{fig:corner_b}
\end{figure}

\begin{figure}
    \centering
    \includegraphics[width=\linewidth]{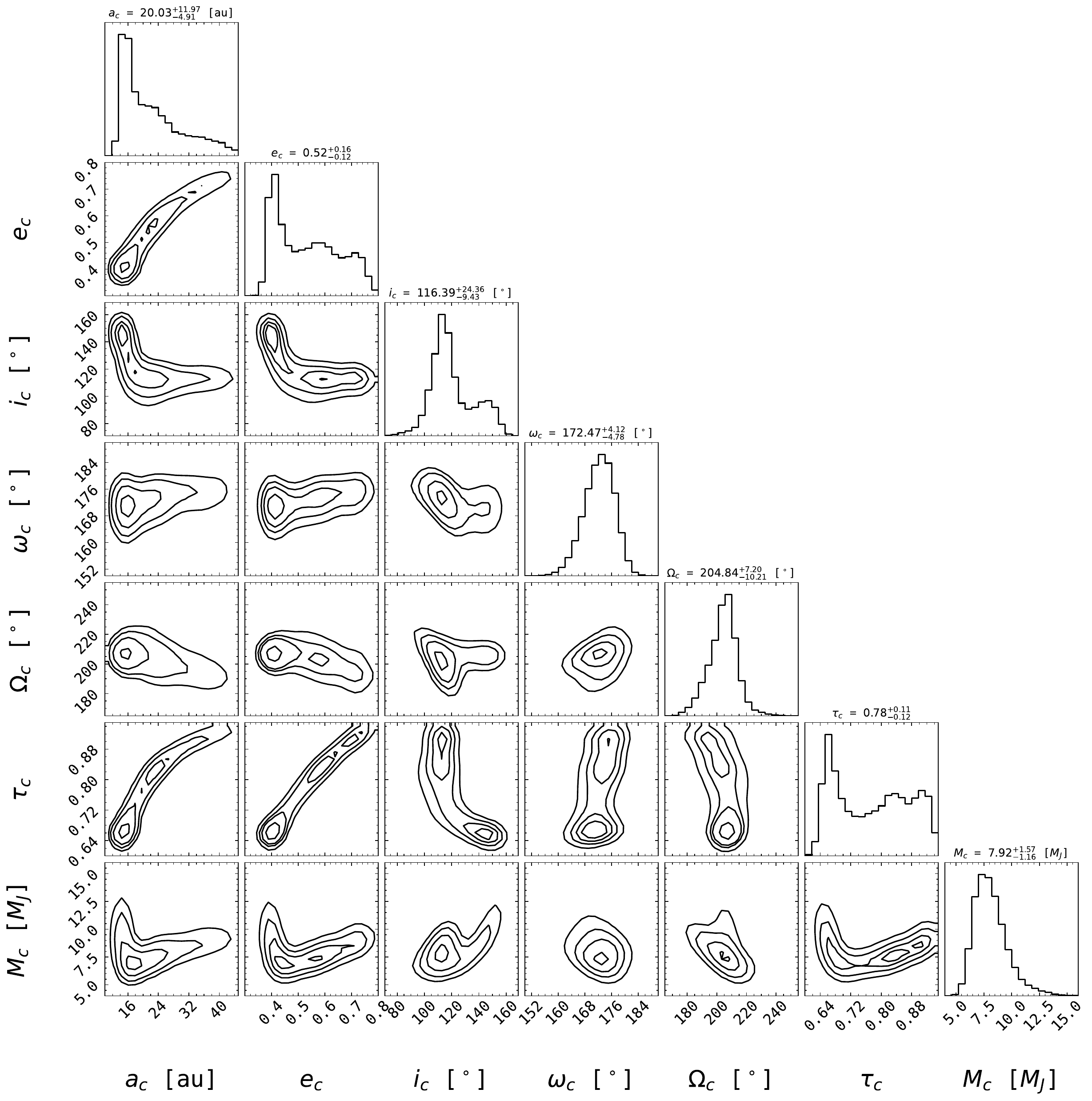}
    \caption{\textbf{Posterior distribution on the orbital elements of 14~Her~c}, from the \texttt{orbitize!} orbit fit. Tails on the distributions are correlated with the rv instrumental jitter terms, due to the fractional phase coverage of the observations from various instruments. An additional epoch of JWST/NIRCam imaging will constrain these parameters dramatically by revising the planet's eccentricity estimate.}
    \label{fig:corner_c}
\end{figure}

For our \texttt{orbitize!} run we use 100 walkers (initialized randomly wihtin the prior space) over 20 temperatures and $2.5{\times}10^{6}$ steps. We discarded the first $2.25{\times}10^{6}$ steps as burn-in and thinned the sampler by a factor of 10, saving only every $10^{th}$ step, resulting in a final posterior distribution with $2.5{\times}10^{6}$ orbits, each composed of 26 parameters. The fit yields two modes in the posterior distribution on $i_b$ and $\Omega_b$, due to a degeneracy in the visual orbit of the inner planet b, while the mode of the parameters describing the orbit of c all have a tail of solutions that are correlated with the RV jitter terms, due to the fractional phase coverage of the planet's orbit by each instrument. Even after the exhaustive MCMC run, the two modes are not equivalently weighted, and the solutions with $i_b=147^{+4.57}_{-9.68}\,^\circ$ are more numerous, at a ratio of about 3 to 1, compared to the solutions with $i_b=32.46^{+7.31}_{-4.25}\,^\circ$. This results in a posterior distribution of mutual inclinations of $\Theta = 32^{+13.6}_{-15.1}\,^{\circ}$ and $\Theta=145.0^{{+15.8}\,\circ}_{-11.1}$, for $i_b$ greater than or less than $90^\circ$, respectively. We adopt this set of posterior parameters and present the median and $1\,\sigma$ confidence intervals of the marginalized posterior distribution in Table \ref{tab:orbitize_results}. Figure \ref{fig:rvs} visualizes the radial velocity measurements compared to 100 randomly drawn posterior models, and Figure \ref{fig:orbs} shows the sky-projected visual orbits for both planets on the same scale. Figures \ref{fig:corner_b} and \ref{fig:corner_c} show the posterior distribution on the orbital elements for 14~Her~b and c respectively, without the nuisance parameters. 

\section{Atmospheric modeling}\label{sec:atm_mod}

We compute custom atmospheric models using \texttt{PICASO} \citep{Batalha2019, Mukherjee2023}, an open-source Python-based framework that solves for one-dimensional pressure–temperature (P–T) profiles in radiative–convective and chemical equilibrium. For cloud modeling, we adopt the \texttt{EddySed} framework outlined in \citep{Ackerman2001}, a method extensively used in previous studies \citep{skemer2016, Morley2015}, as well as in model grids such as those from \citep{Saumon2008} and the Sonora Diamondback suite \citep{Morley2024}. Within \texttt{PICASO}, we incorporate an updated cloud treatment using \texttt{Virga} \citep{virga}, a Python-based implementation of the \citep{Ackerman2001} model. The mini-grid spans effective temperatures of $T_{\rm eff}$ = [275, 300, 325 K], surface gravity of log(g) = 4.25 (cgs), eddy diffusion parameter of $K_{\rm zz}$ = [10$^2$, 10$^7$, 10$^9$ cm$^2$ s$^{-1}$], a metallicity of [M/H] = +0.5 (relative to the sun), and a solar C/O ratio. Cloudy models have an $f_{\rm sed}$ = 8, with H$_2$O as the only condensing species. These models are part of a larger grid that will be released in the future, extending the Sonora Bobcat \citep{Marley2021}, Sonora Diamondback \citep{Morley2024}, and Sonora Elf Owl \citep{Mukherjee2024EO}.

Our custom models differ from the Sonora Elf Owl models in three key aspects of atmospheric chemistry. First, we adopt an updated treatment of CO$_2$. Previous studies have shown that Sonora Elf Owl underestimates CO$_2$ abundances compared to JWST observations of cool substellar objects \citep{Beiler2024}. To address this imbalance, we implement a CO$_2$ kinetic model based on the chemical kinetics described in \citep{Zahnle2014}, following recent investigations into its impact \citep{Wogan2023, Mukherjee2024Photochem}. This update leads to a stronger CO$_2$ absorption feature near 4.2 $\mu$m. Second, unlike Sonora Elf Owl, our models incorporate rainout chemistry for condensates such as H$_2$O, even in cloud-free atmospheres, following the chemical rainout treatment in the Sonora Bobcat models. Finally, we exclude PH$_3$, as its abundance was found to be overestimated in Elf Owl models compared to cool brown dwarf observations \citep{Beiler2024}.

\section{Potential for von Zeipel-Lidov-Kozai oscillations}

The orbital structure of the 14 Her system suggests the possibility that the system might be experiencing von Zeipel-Lidov-Kozai (ZLK) oscillations, characterized by coupled high-amplitude eccentricity and inclination oscillations in misaligned, hierarchical three-body systems between the ZLK critical angles $39.2^{\circ} < \Theta < 140.8^{\circ}$. However, the very concept of ZLK is ill-defined for 14 Her-like systems. The prototypical ZLK effect is defined in the limit of a massless test particle and a massive perturber on a circular orbit. The further a system strays from this regime, the more the system's secular evolution deviates from traditional ZLK behavior \citep{2016ARA&A..54..441N}. In particular, the relatively equal planetary masses in the 14 Her system make classifying a particular system evolution as ZLK behavior not just difficult, but arguably meaningless. Regardless of the specific class of oscillation, we emphasize that the system at present is undoubtedly undergoing extreme mutual dynamical interactions.

About $30\%$ of the inferred distribution of the mutual inclination $\Theta$ lies between the ZLK critical angles \citep{vonZeipel1910, Lidov1962, 1962AJ.....67R.579K}. The ZLK mechanism is thought to have widespread consequences for planetary systems, including elongating the orbits of comets in our Solar System, as well as triggering high eccentricity migration of planets to become hot Jupiters~\citep{2011Natur.473..187N}. While a number of extrasolar systems are believed to have experienced ZLK oscillations in the past \citep{wu2003}, 14 Her is the first instance in which an ongoing ZLK mechanism may be observed on a mature system with long-period exoplanets.



\bibliography{14her_refs}{}

\begin{thebibliography}{}
\expandafter\ifx\csname natexlab\endcsname\relax\def\natexlab#1{#1}\fi
\providecommand{\url}[1]{\href{#1}{#1}}
\providecommand{\dodoi}[1]{doi:~\href{http://doi.org/#1}{\nolinkurl{#1}}}
\providecommand{\doeprint}[1]{\href{http://ascl.net/#1}{\nolinkurl{http://ascl.net/#1}}}
\providecommand{\doarXiv}[1]{\href{https://arxiv.org/abs/#1}{\nolinkurl{https://arxiv.org/abs/#1}}}

\bibitem[{A.~S. {Ackerman} \& M.~S. {Marley}(2001){Ackerman} \&
  {Marley}}]{Ackerman2001}
{Ackerman}, A.~S., \& {Marley}, M.~S. 2001, \bibinfo{title}{{Precipitating
  Condensation Clouds in Substellar Atmospheres},} \apj, 556, 872,
  \dodoi{10.1086/321540}

\bibitem[{J.~I. {Adams Redai} {et~al.}(2023){Adams Redai}, {Follette}, {Wang},
  {Leonard}, {Balmer}, {Close}, {Dacus}, {Males}, {Morzinski}, {Palmo},
  {Pueyo}, {Spiro}, {Treiber}, {Ward-Duong}, \& {Watson}}]{AdamsRedai2023}
{Adams Redai}, J.~I., {Follette}, K.~B., {Wang}, J., {et~al.} 2023,
  \bibinfo{title}{{The Giant Accreting Protoplanet Survey (GAPlanetS):
  Optimization Techniques for Robust Detections of Protoplanets},} \aj, 165,
  57, \dodoi{10.3847/1538-3881/aca60d}

\bibitem[{F. {Allard} {et~al.}(2011){Allard}, {Homeier}, \&
  {Freytag}}]{Allard2011}
{Allard}, F., {Homeier}, D., \& {Freytag}, B. 2011, in Astronomical Society of
  the Pacific Conference Series, Vol. 448, 16th Cambridge Workshop on Cool
  Stars, Stellar Systems, and the Sun, ed. C.~{Johns-Krull}, M.~K. {Browning},
  \& A.~A. {West}, 91, \dodoi{10.48550/arXiv.1011.5405}

\bibitem[{Q. {An} {et~al.}(2025){An}, {Lu}, {Brandt}, {Brandt}, \&
  {Li}}]{An2025}
{An}, Q., {Lu}, T., {Brandt}, G.~M., {Brandt}, T.~D., \& {Li}, G. 2025,
  \bibinfo{title}{{Significant Mutual Inclinations Between the Stellar Spin and
  the Orbits of Both Planets in the HAT-P-11 System},} \aj, 169, 22,
  \dodoi{10.3847/1538-3881/ad90b4}

\bibitem[{W.~O. Balmer(2025)Balmer}]{reduction_doi}
Balmer, W.~O. 2025, \bibinfo{title}{wbalmer/a\_herculean\_detection: 14
  Herculis c JWST coronagraphic imaging,} Zenodo,
  \dodoi{10.5281/ZENODO.15483953}

\bibitem[{W.~O. {Balmer} {et~al.}(2025){Balmer}, {Kammerer}, {Pueyo}, {Perrin},
  {Girard}, {Leisenring}, {Lawson}, {Dennen}, {van der Marel}, {Beichman},
  {Bryden}, {Llop-Sayson}, {Valenti}, {Lothringer}, {Lewis}, {M{\^a}lin},
  {Rebollido}, {Rickman}, {Hoch}, {Soummer}, {Clampin}, \&
  {Mountain}}]{Balmer2025b}
{Balmer}, W.~O., {Kammerer}, J., {Pueyo}, L., {et~al.} 2025,
  \bibinfo{title}{{JWST-TST High Contrast: Living on the Wedge, or, NIRCam Bar
  Coronagraphy Reveals CO$_{2}$ in the HR 8799 and 51 Eri Exoplanets'
  Atmospheres},} \aj, 169, 209, \dodoi{10.3847/1538-3881/adb1c6}

\bibitem[{D.~C. {Bardalez Gagliuffi} {et~al.}(2021){Bardalez Gagliuffi},
  {Faherty}, {Li}, {Brandt}, {Williams}, {Brandt}, \&
  {Gelino}}]{2021ApJ...922L..43B}
{Bardalez Gagliuffi}, D.~C., {Faherty}, J.~K., {Li}, Y., {et~al.} 2021,
  \bibinfo{title}{{14 Her: A Likely Case of Planet-Planet Scattering},} \apjl,
  922, L43, \dodoi{10.3847/2041-8213/ac382c}

\bibitem[{N. Batalha {et~al.}(2020)Batalha, caoimherooney11, \&
  sagnickm}]{virga}
Batalha, N., caoimherooney11, \& sagnickm. 2020,
  \bibinfo{title}{natashabatalha/virga: Initial Release,}, v0.0 Zenodo,
  \dodoi{10.5281/zenodo.3759888}

\bibitem[{N.~E. {Batalha} {et~al.}(2019){Batalha}, {Marley}, {Lewis}, \&
  {Fortney}}]{Batalha2019}
{Batalha}, N.~E., {Marley}, M.~S., {Lewis}, N.~K., \& {Fortney}, J.~J. 2019,
  \bibinfo{title}{{Exoplanet Reflected-light Spectroscopy with PICASO},} \apj,
  878, 70, \dodoi{10.3847/1538-4357/ab1b51}

\bibitem[{S.~A. {Beiler} {et~al.}(2023){Beiler}, {Cushing}, {Kirkpatrick},
  {Schneider}, {Mukherjee}, \& {Marley}}]{2023ApJ...951L..48B}
{Beiler}, S.~A., {Cushing}, M.~C., {Kirkpatrick}, J.~D., {et~al.} 2023,
  \bibinfo{title}{{The First JWST Spectral Energy Distribution of a Y Dwarf},}
  \apjl, 951, L48, \dodoi{10.3847/2041-8213/ace32c}

\bibitem[{S.~A. {Beiler} {et~al.}(2024{\natexlab{a}}){Beiler}, {Cushing},
  {Kirkpatrick}, {Schneider}, {Mukherjee}, {Marley}, {Marocco}, \&
  {Smart}}]{2024ApJ...973..107B}
{Beiler}, S.~A., {Cushing}, M.~C., {Kirkpatrick}, J.~D., {et~al.}
  2024{\natexlab{a}}, \bibinfo{title}{{Precise Bolometric Luminosities and
  Effective Temperatures of 23 Late-T and Y Dwarfs Obtained with JWST},} \apj,
  973, 107, \dodoi{10.3847/1538-4357/ad6301}

\bibitem[{S.~A. {Beiler} {et~al.}(2024{\natexlab{b}}){Beiler}, {Mukherjee},
  {Cushing}, {Kirkpatrick}, {Schneider}, {Kothari}, {Marley}, \&
  {Visscher}}]{Beiler2024}
{Beiler}, S.~A., {Mukherjee}, S., {Cushing}, M.~C., {et~al.}
  2024{\natexlab{b}}, \bibinfo{title}{{A Tale of Two Molecules: The
  Underprediction of CO$_{2}$ and Overprediction of PH$_{3}$ in Late T and Y
  Dwarf Atmospheric Models},} \apj, 973, 60, \dodoi{10.3847/1538-4357/ad6759}

\bibitem[{G.~F. {Benedict} {et~al.}(2023){Benedict}, {McArthur}, {Nelan}, \&
  {Bean}}]{2023AJ....166...27B}
{Benedict}, G.~F., {McArthur}, B.~E., {Nelan}, E.~P., \& {Bean}, J.~L. 2023,
  \bibinfo{title}{{The 14 Her Planetary System: Companion Masses and
  Architecture from Radial Velocities and Astrometry},} \aj, 166, 27,
  \dodoi{10.3847/1538-3881/acd93a}

\bibitem[{S. {Blunt} {et~al.}(2020){Blunt}, {Wang}, {Angelo}, {Ngo}, {Cody},
  {De Rosa}, {Graham}, {Hirsch}, {Nagpal}, {Nielsen}, {Pearce}, {Rice}, \&
  {Tejada}}]{2020AJ....159...89B}
{Blunt}, S., {Wang}, J.~J., {Angelo}, I., {et~al.} 2020,
  \bibinfo{title}{{orbitize!: A Comprehensive Orbit-fitting Software Package
  for the High-contrast Imaging Community},} \aj, 159, 89,
  \dodoi{10.3847/1538-3881/ab6663}

\bibitem[{E. {Bogat} {et~al.}(2025){Bogat}, {Schlieder}, {Lawson}, {Li},
  {Leisenring}, {Meyer}, {Balmer}, {Barclay}, {Beichman}, {Bryden},
  {Calissendorff}, {Carter}, {De Furio}, {Girard}, {Greene}, {Groff},
  {Kammerer}, {Llop-Sayson}, {McElwain}, {Rieke}, \& {Ygouf}}]{Bogat2025}
{Bogat}, E., {Schlieder}, J.~E., {Lawson}, K.~D., {et~al.} 2025,
  \bibinfo{title}{{Probing the Outskirts of M Dwarf Planetary Systems with a
  Cycle 1 JWST NIRCam Coronagraphy Survey},} arXiv e-prints, arXiv:2504.11659,
  \dodoi{10.48550/arXiv.2504.11659}

\bibitem[{V. {Bourrier} {et~al.}(2021){Bourrier}, {Lovis}, {Cretignier},
  {Allart}, {Dumusque}, {Delisle}, {Deline}, {Sousa}, {Adibekyan}, {Alibert},
  {Barros}, {Borsa}, {Cristiani}, {Demangeon}, {Ehrenreich}, {Figueira},
  {Gonz{\'a}lez Hern{\'a}ndez}, {Lendl}, {Lillo-Box}, {Lo Curto}, {Di
  Marcantonio}, {Martins}, {M{\'e}gevand}, {Mehner}, {Micela}, {Molaro},
  {Oshagh}, {Palle}, {Pepe}, {Poretti}, {Rebolo}, {Santos}, {Scandariato},
  {Seidel}, {Sozzetti}, {Su{\'a}rez Mascare{\~n}o}, \& {Zapatero
  Osorio}}]{2021A&A...654A.152B}
{Bourrier}, V., {Lovis}, C., {Cretignier}, M., {et~al.} 2021,
  \bibinfo{title}{{The Rossiter-McLaughlin effect revolutions: an ultra-short
  period planet and a warm mini-Neptune on perpendicular orbits},} \aap, 654,
  A152, \dodoi{10.1051/0004-6361/202141527}

\bibitem[{R. {Bowens-Rubin} {et~al.}(2025){Bowens-Rubin}, {Mang}, {Limbach},
  {Carter}, {Stevenson}, {Wagner}, {Strampelli}, {Morley}, {Kennedy},
  {Matthews}, {Vanderburg}, \& {Salama}}]{Bowens-Rubin2025}
{Bowens-Rubin}, R., {Mang}, J., {Limbach}, M.~A., {et~al.} 2025,
  \bibinfo{title}{{NIRCam yells at cloud: JWST MIRI imaging can directly detect
  exoplanets of the same temperature, mass, age, and orbital separation as
  Saturn and Jupiter},} arXiv e-prints, arXiv:2505.15995,
  \dodoi{10.48550/arXiv.2505.15995}

\bibitem[{B.~P. {Bowler}(2016){Bowler}}]{2016PASP..128j2001B}
{Bowler}, B.~P. 2016, \bibinfo{title}{{Imaging Extrasolar Giant Planets},}
  \pasp, 128, 102001, \dodoi{10.1088/1538-3873/128/968/102001}

\bibitem[{B.~P. {Bowler} {et~al.}(2020){Bowler}, {Blunt}, \&
  {Nielsen}}]{2020AJ....159...63B}
{Bowler}, B.~P., {Blunt}, S.~C., \& {Nielsen}, E.~L. 2020,
  \bibinfo{title}{{Population-level Eccentricity Distributions of Imaged
  Exoplanets and Brown Dwarf Companions: Dynamical Evidence for Distinct
  Formation Channels},} \aj, 159, 63, \dodoi{10.3847/1538-3881/ab5b11}

\bibitem[{T.~D. {Brandt}(2018){Brandt}}]{2018ApJS..239...31B}
{Brandt}, T.~D. 2018, \bibinfo{title}{{The Hipparcos{\ndash}Gaia Catalog of
  Accelerations},} \apjs, 239, 31, \dodoi{10.3847/1538-4365/aaec06}

\bibitem[{T.~D. {Brandt}(2024{\natexlab{a}}){Brandt}}]{Brandt2024a}
{Brandt}, T.~D. 2024{\natexlab{a}}, \bibinfo{title}{{Optimal Fitting and
  Debiasing for Detectors Read Out Up-the-Ramp},} \pasp, 136, 045004,
  \dodoi{10.1088/1538-3873/ad38d9}

\bibitem[{T.~D. {Brandt}(2024{\natexlab{b}}){Brandt}}]{Brandt2024b}
{Brandt}, T.~D. 2024{\natexlab{b}}, \bibinfo{title}{{Likelihood-based Jump
  Detection and Cosmic Ray Rejection for Detectors Read Out Up-the-ramp},}
  \pasp, 136, 045005, \dodoi{10.1088/1538-3873/ad38da}

\bibitem[{T.~D. {Brandt} {et~al.}(2021){Brandt}, {Dupuy}, {Li}, {Brandt},
  {Zeng}, {Michalik}, {Bardalez Gagliuffi}, \&
  {Raposo-Pulido}}]{2021AJ....162..186B}
{Brandt}, T.~D., {Dupuy}, T.~J., {Li}, Y., {et~al.} 2021,
  \bibinfo{title}{{orvara: An Efficient Code to Fit Orbits Using Radial
  Velocity, Absolute, and/or Relative Astrometry},} \aj, 162, 186,
  \dodoi{10.3847/1538-3881/ac042e}

\bibitem[{J. {Buchner} {et~al.}(2014){Buchner}, {Georgakakis}, {Nandra}, {Hsu},
  {Rangel}, {Brightman}, {Merloni}, {Salvato}, {Donley}, \&
  {Kocevski}}]{Buchner2014}
{Buchner}, J., {Georgakakis}, A., {Nandra}, K., {et~al.} 2014,
  \bibinfo{title}{{X-ray spectral modelling of the AGN obscuring region in the
  CDFS: Bayesian model selection and catalogue},} \aap, 564, A125,
  \dodoi{10.1051/0004-6361/201322971}

\bibitem[{H. {Bushouse} {et~al.}(2023){Bushouse}, {Eisenhamer}, {Dencheva},
  {Davies}, {Greenfield}, {Morrison}, {Hodge}, {Simon}, {Grumm}, {Droettboom},
  {Slavich}, {Sosey}, {Pauly}, {Miller}, {Jedrzejewski}, {Hack}, {Davis},
  {Crawford}, {Law}, {Gordon}, {Regan}, {Cara}, {MacDonald}, {Bradley},
  {Shanahan}, {Jamieson}, {Teodoro}, {Williams}, \&
  {Pena-Guerrero}}]{Bushouse2023}
{Bushouse}, H., {Eisenhamer}, J., {Dencheva}, N., {et~al.} 2023,
  \bibinfo{title}{{JWST Calibration Pipeline},}, 1.12.1, Zenodo Zenodo,
  \dodoi{10.5281/zenodo.6984365}

\bibitem[{J.~C. {Carson} {et~al.}(2009){Carson}, {Hiner}, {Villar}, {Blaschak},
  {Rudolph}, \& {Stapelfeldt}}]{Carson2009}
{Carson}, J.~C., {Hiner}, K.~D., {Villar}, Gregorio~G., I., {et~al.} 2009,
  \bibinfo{title}{{A Distance-Limited Imaging Survey of Substellar Companions
  to Solar Neighborhood Stars},} \aj, 137, 218,
  \dodoi{10.1088/0004-6256/137/1/218}

\bibitem[{A.~L. {Carter} {et~al.}(2023){Carter}, {Hinkley}, {Kammerer},
  {Skemer}, {Biller}, {Leisenring}, {Millar-Blanchaer}, {Petrus}, {Stone},
  {Ward-Duong}, {Wang}, {Girard}, {Hines}, {Perrin}, {Pueyo}, {Balmer},
  {Bonavita}, {Bonnefoy}, {Chauvin}, {Choquet}, {Christiaens}, {Danielski},
  {Kennedy}, {Matthews}, {Miles}, {Patapis}, {Ray}, {Rickman}, {Sallum},
  {Stapelfeldt}, {Whiteford}, {Zhou}, {Absil}, {Boccaletti}, {Booth}, {Bowler},
  {Chen}, {Currie}, {Fortney}, {Grady}, {Greebaum}, {Henning}, {Hoch},
  {Janson}, {Kalas}, {Kenworthy}, {Kervella}, {Kraus}, {Lagage}, {Liu},
  {Macintosh}, {Marino}, {Marley}, {Marois}, {Matthews}, {Mawet}, {McElwain},
  {Metchev}, {Meyer}, {Molliere}, {Moran}, {Morley}, {Mukherjee}, {Pantin},
  {Quirrenbach}, {Rebollido}, {Ren}, {Schneider}, {Vasist}, {Worthen}, {Wyatt},
  {Briesemeister}, {Bryan}, {Calissendorff}, {Cantalloube}, {Cugno}, {De
  Furio}, {Dupuy}, {Factor}, {Faherty}, {Fitzgerald}, {Franson}, {Gonzales},
  {Hood}, {Howe}, {Kuzuhara}, {Lagrange}, {Lawson}, {Lazzoni}, {Lew}, {Liu},
  {Llop-Sayson}, {Lloyd}, {Martinez}, {Mazoyer}, {Palma-Bifani}, {Quanz},
  {Redai}, {Samland}, {Schlieder}, {Tamura}, {Tan}, {Uyama}, {Vigan}, {Vos},
  {Wagner}, {Wolff}, {Ygouf}, {Zhang}, {Zhang}, \& {Zhang}}]{Carter2023}
{Carter}, A.~L., {Hinkley}, S., {Kammerer}, J., {et~al.} 2023,
  \bibinfo{title}{{The JWST Early Release Science Program for Direct
  Observations of Exoplanetary Systems I: High-contrast Imaging of the
  Exoplanet HIP 65426 b from 2 to 16 {\ensuremath{\mu}}m},} \apjl, 951, L20,
  \dodoi{10.3847/2041-8213/acd93e}

\bibitem[{I. {Czekala} {et~al.}(2019){Czekala}, {Chiang}, {Andrews}, {Jensen},
  {Torres}, {Wilner}, {Stassun}, \& {Macintosh}}]{2019ApJ...883...22C}
{Czekala}, I., {Chiang}, E., {Andrews}, S.~M., {et~al.} 2019,
  \bibinfo{title}{{The Degree of Alignment between Circumbinary Disks and Their
  Binary Hosts},} \apj, 883, 22, \dodoi{10.3847/1538-4357/ab287b}

\bibitem[{R.~J. {De Rosa} {et~al.}(2020){De Rosa}, {Dawson}, \&
  {Nielsen}}]{2020A&A...640A..73D}
{De Rosa}, R.~J., {Dawson}, R., \& {Nielsen}, E.~L. 2020, \bibinfo{title}{{A
  significant mutual inclination between the planets within the
  {\ensuremath{\pi}} Mensae system},} \aap, 640, A73,
  \dodoi{10.1051/0004-6361/202038496}

\bibitem[{R. {Deitrick} {et~al.}(2015){Deitrick}, {Barnes}, {McArthur},
  {Quinn}, {Luger}, {Antonsen}, \& {Benedict}}]{2015ApJ...798...46D}
{Deitrick}, R., {Barnes}, R., {McArthur}, B., {et~al.} 2015,
  \bibinfo{title}{{The Three-dimensional Architecture of the
  {\ensuremath{\upsilon}} Andromedae Planetary System},} \apj, 798, 46,
  \dodoi{10.1088/0004-637X/798/1/46}

\bibitem[{S. {Durkan} {et~al.}(2016){Durkan}, {Janson}, \&
  {Carson}}]{Durkan2016}
{Durkan}, S., {Janson}, M., \& {Carson}, J.~C. 2016, \bibinfo{title}{{High
  Contrast Imaging with Spitzer: Constraining the Frequency of Giant Planets
  out to 1000 au Separations},} \apj, 824, 58,
  \dodoi{10.3847/0004-637X/824/1/58}

\bibitem[{P.~P. {Eggleton} \& L. {Kiseleva-Eggleton}(2001){Eggleton} \&
  {Kiseleva-Eggleton}}]{2001ApJ...562.1012E}
{Eggleton}, P.~P., \& {Kiseleva-Eggleton}, L. 2001, \bibinfo{title}{{Orbital
  Evolution in Binary and Triple Stars, with an Application to SS Lacertae},}
  \apj, 562, 1012, \dodoi{10.1086/323843}

\bibitem[{D.~J. {Eisenstein} {et~al.}(2023){Eisenstein}, {Johnson},
  {Robertson}, {Tacchella}, {Hainline}, {Jakobsen}, {Maiolino}, {Bonaventura},
  {Bunker}, {Cameron}, {Cargile}, {Curtis-Lake}, {Hausen}, {Pusk{\'a}s},
  {Rieke}, {Sun}, {Willmer}, {Willott}, {Alberts}, {Arribas}, {Baker}, {Baum},
  {Bhatawdekar}, {Carniani}, {Charlot}, {Chen}, {Chevallard}, {Curti},
  {DeCoursey}, {D'Eugenio}, {de Graaff}, {Egami}, {Helton}, {Ji}, {Jones},
  {Kumari}, {L{\"u}tzgendorf}, {Laseter}, {Looser}, {Lyu}, {Maseda}, {Nelson},
  {Parlanti}, {Rauscher}, {Rawle}, {Rieke}, {Rix}, {Rujopakarn}, {Sandles},
  {Saxena}, {Scholtz}, {Sharpe}, {Shivaei}, {Simmonds}, {Smit}, {Topping},
  {{\"U}bler}, {Venturi}, {Williams}, {Witstok}, \& {Woodrum}}]{Eisenstein2023}
{Eisenstein}, D.~J., {Johnson}, B.~D., {Robertson}, B., {et~al.} 2023,
  \bibinfo{title}{{The JADES Origins Field: A New JWST Deep Field in the JADES
  Second NIRCam Data Release},} arXiv e-prints, arXiv:2310.12340,
  \dodoi{10.48550/arXiv.2310.12340}

\bibitem[{D. {Fabrycky} \& S. {Tremaine}(2007){Fabrycky} \&
  {Tremaine}}]{2007ApJ...669.1298F}
{Fabrycky}, D., \& {Tremaine}, S. 2007, \bibinfo{title}{{Shrinking Binary and
  Planetary Orbits by Kozai Cycles with Tidal Friction},} \apj, 669, 1298,
  \dodoi{10.1086/521702}

\bibitem[{F. {Feng} {et~al.}(2024){Feng}, {Xiao}, {Jones}, {Jenkins}, {Pena},
  \& {Sun}}]{2024arXiv241214542F}
{Feng}, F., {Xiao}, G.-Y., {Jones}, H. R.~A., {et~al.} 2024,
  \bibinfo{title}{{Lessons learned from the detection of wide companions by
  radial velocity and astrometry},} arXiv e-prints, arXiv:2412.14542,
  \dodoi{10.48550/arXiv.2412.14542}

\bibitem[{F. {Feng} {et~al.}(2022){Feng}, {Butler}, {Vogt}, {Clement},
  {Tinney}, {Cui}, {Aizawa}, {Jones}, {Bailey}, {Burt}, {Carter}, {Crane},
  {Flammini Dotti}, {Holden}, {Ma}, {Ogihara}, {Oppenheimer}, {O'Toole},
  {Shectman}, {Wittenmyer}, {Wang}, {Wright}, \& {Xuan}}]{2022ApJS..262...21F}
{Feng}, F., {Butler}, R.~P., {Vogt}, S.~S., {et~al.} 2022, \bibinfo{title}{{3D
  Selection of 167 Substellar Companions to Nearby Stars},} \apjs, 262, 21,
  \dodoi{10.3847/1538-4365/ac7e57}

\bibitem[{F. {Feroz} {et~al.}(2009){Feroz}, {Hobson}, \& {Bridges}}]{Feroz2009}
{Feroz}, F., {Hobson}, M.~P., \& {Bridges}, M. 2009,
  \bibinfo{title}{{MULTINEST: an efficient and robust Bayesian inference tool
  for cosmology and particle physics},} \mnras, 398, 1601,
  \dodoi{10.1111/j.1365-2966.2009.14548.x}

\bibitem[{K.~B. {Follette}(2023){Follette}}]{Follette2023}
{Follette}, K.~B. 2023, \bibinfo{title}{{An Introduction to High Contrast
  Differential Imaging of Exoplanets and Disks},} \pasp, 135, 093001,
  \dodoi{10.1088/1538-3873/aceb31}

\bibitem[{D. {Foreman-Mackey} {et~al.}(2013){Foreman-Mackey}, {Hogg}, {Lang},
  \& {Goodman}}]{2013PASP..125..306F}
{Foreman-Mackey}, D., {Hogg}, D.~W., {Lang}, D., \& {Goodman}, J. 2013,
  \bibinfo{title}{{emcee: The MCMC Hammer},} \pasp, 125, 306,
  \dodoi{10.1086/670067}

\bibitem[{K. {Franson} {et~al.}(2024{\natexlab{a}}){Franson}, {Balmer},
  {Bowler}, {Pueyo}, {Zhou}, {Rickman}, {Zhang}, {Mukherjee}, {Pearce},
  {Bardalez Gagliuffi}, {Biddle}, {Brandt}, {Bowens-Rubin}, {Crepp},
  {Davidson}, {Faherty}, {Ginski}, {Horch}, {Morgan}, {Morley}, {Perrin},
  {Sanghi}, {Salama}, {Theissen}, {Tran}, \& {Wolf}}]{2024ApJ...974L..11F}
{Franson}, K., {Balmer}, W.~O., {Bowler}, B.~P., {et~al.} 2024{\natexlab{a}},
  \bibinfo{title}{{JWST/NIRCam 4{\textendash}5 {\ensuremath{\mu}}m Imaging of
  the Giant Planet AF Lep b},} \apjl, 974, L11,
  \dodoi{10.3847/2041-8213/ad736a}

\bibitem[{K. {Franson} {et~al.}(2024{\natexlab{b}}){Franson}, {Balmer},
  {Bowler}, {Pueyo}, {Zhou}, {Rickman}, {Zhang}, {Mukherjee}, {Pearce},
  {Bardalez Gagliuffi}, {Biddle}, {Brandt}, {Bowens-Rubin}, {Crepp},
  {Davidson}, {Faherty}, {Ginski}, {Horch}, {Morgan}, {Morley}, {Perrin},
  {Sanghi}, {Salama}, {Theissen}, {Tran}, \& {Wolf}}]{Franson2024}
{Franson}, K., {Balmer}, W.~O., {Bowler}, B.~P., {et~al.} 2024{\natexlab{b}},
  \bibinfo{title}{{JWST/NIRCam 4{\textendash}5 {\ensuremath{\mu}}m Imaging of
  the Giant Planet AF Lep b},} \apjl, 974, L11,
  \dodoi{10.3847/2041-8213/ad736a}

\bibitem[{ {Gaia Collaboration} {et~al.}(2021){Gaia Collaboration}, {Brown},
  {Vallenari}, {Prusti}, {de Bruijne}, {Babusiaux}, {Biermann}, {Creevey},
  {Evans}, {Eyer}, {Hutton}, {Jansen}, {Jordi}, {Klioner}, {Lammers},
  {Lindegren}, {Luri}, {Mignard}, {Panem}, {Pourbaix}, {Randich}, {Sartoretti},
  {Soubiran}, {Walton}, {Arenou}, {Bailer-Jones}, {Bastian}, {Cropper},
  {Drimmel}, {Katz}, {Lattanzi}, {van Leeuwen}, {Bakker}, {Cacciari},
  {Casta{\~n}eda}, {De Angeli}, {Ducourant}, {Fabricius}, {Fouesneau},
  {Fr{\'e}mat}, {Guerra}, {Guerrier}, {Guiraud}, {Jean-Antoine Piccolo},
  {Masana}, {Messineo}, {Mowlavi}, {Nicolas}, {Nienartowicz}, {Pailler},
  {Panuzzo}, {Riclet}, {Roux}, {Seabroke}, {Sordo}, {Tanga}, {Th{\'e}venin},
  {Gracia-Abril}, {Portell}, {Teyssier}, {Altmann}, {Andrae}, {Bellas-Velidis},
  {Benson}, {Berthier}, {Blomme}, {Brugaletta}, {Burgess}, {Busso}, {Carry},
  {Cellino}, {Cheek}, {Clementini}, {Damerdji}, {Davidson}, {Delchambre},
  {Dell'Oro}, {Fern{\'a}ndez-Hern{\'a}ndez}, {Galluccio}, {Garc{\'\i}a-Lario},
  {Garcia-Reinaldos}, {Gonz{\'a}lez-N{\'u}{\~n}ez}, {Gosset}, {Haigron},
  {Halbwachs}, {Hambly}, {Harrison}, {Hatzidimitriou}, {Heiter},
  {Hern{\'a}ndez}, {Hestroffer}, {Hodgkin}, {Holl}, {Jan{\ss}en}, {Jevardat de
  Fombelle}, {Jordan}, {Krone-Martins}, {Lanzafame}, {L{\"o}ffler}, {Lorca},
  {Manteiga}, {Marchal}, {Marrese}, {Moitinho}, {Mora}, {Muinonen}, {Osborne},
  {Pancino}, {Pauwels}, {Petit}, {Recio-Blanco}, {Richards}, {Riello},
  {Rimoldini}, {Robin}, {Roegiers}, {Rybizki}, {Sarro}, {Siopis}, {Smith},
  {Sozzetti}, {Ulla}, {Utrilla}, {van Leeuwen}, {van Reeven}, {Abbas}, {Abreu
  Aramburu}, {Accart}, {Aerts}, {Aguado}, {Ajaj}, {Altavilla}, {{\'A}lvarez},
  {{\'A}lvarez Cid-Fuentes}, {Alves}, {Anderson}, {Anglada Varela}, {Antoja},
  {Audard}, {Baines}, {Baker}, {Balaguer-N{\'u}{\~n}ez}, {Balbinot}, {Balog},
  {Barache}, {Barbato}, {Barros}, {Barstow}, {Bartolom{\'e}}, {Bassilana},
  {Bauchet}, {Baudesson-Stella}, {Becciani}, {Bellazzini}, {Bernet}, {Bertone},
  {Bianchi}, {Blanco-Cuaresma}, {Boch}, {Bombrun}, {Bossini}, {Bouquillon},
  {Bragaglia}, {Bramante}, {Breedt}, {Bressan}, {Brouillet}, {Bucciarelli},
  {Burlacu}, {Busonero}, {Butkevich}, {Buzzi}, {Caffau}, {Cancelliere},
  {C{\'a}novas}, {Cantat-Gaudin}, {Carballo}, {Carlucci}, {Carnerero},
  {Carrasco}, {Casamiquela}, {Castellani}, {Castro-Ginard}, {Castro Sampol},
  {Chaoul}, {Charlot}, {Chemin}, {Chiavassa}, {Cioni}, {Comoretto}, {Cooper},
  {Cornez}, {Cowell}, {Crifo}, {Crosta}, {Crowley}, {Dafonte}, {Dapergolas},
  {David}, {David}, {de Laverny}, {De Luise}, {De March}, {De Ridder}, {de
  Souza}, {de Teodoro}, {de Torres}, {del Peloso}, {del Pozo}, {Delbo},
  {Delgado}, {Delgado}, {Delisle}, {Di Matteo}, {Diakite}, {Diener},
  {Distefano}, {Dolding}, {Eappachen}, {Edvardsson}, {Enke}, {Esquej}, {Fabre},
  {Fabrizio}, {Faigler}, {Fedorets}, {Fernique}, {Fienga}, {Figueras},
  {Fouron}, {Fragkoudi}, {Fraile}, {Franke}, {Gai}, {Garabato},
  {Garcia-Gutierrez}, {Garc{\'\i}a-Torres}, {Garofalo}, {Gavras}, {Gerlach},
  {Geyer}, {Giacobbe}, {Gilmore}, {Girona}, {Giuffrida}, {Gomel}, {Gomez},
  {Gonzalez-Santamaria}, {Gonz{\'a}lez-Vidal}, {Granvik},
  {Guti{\'e}rrez-S{\'a}nchez}, {Guy}, {Hauser}, {Haywood}, {Helmi}, {Hidalgo},
  {Hilger}, {H{\l}adczuk}, {Hobbs}, {Holland}, {Huckle}, {Jasniewicz},
  {Jonker}, {Juaristi Campillo}, {Julbe}, {Karbevska}, {Kervella}, {Khanna},
  {Kochoska}, {Kontizas}, {Kordopatis}, {Korn}, {Kostrzewa-Rutkowska},
  {Kruszy{\'n}ska}, {Lambert}, {Lanza}, {Lasne}, {Le Campion}, {Le Fustec},
  {Lebreton}, {Lebzelter}, {Leccia}, {Leclerc}, {Lecoeur-Taibi}, {Liao},
  {Licata}, {Lindstr{\o}m}, {Lister}, {Livanou}, {Lobel}, {Madrero Pardo},
  {Managau}, {Mann}, {Marchant}, {Marconi}, {Marcos Santos}, {Marinoni},
  {Marocco}, {Marshall}, {Martin Polo}, {Mart{\'\i}n-Fleitas}, {Masip},
  {Massari}, {Mastrobuono-Battisti}, {Mazeh}, {McMillan}, {Messina},
  {Michalik}, {Millar}, {Mints}, {Molina}, {Molinaro}, {Moln{\'a}r},
  {Montegriffo}, {Mor}, {Morbidelli}, {Morel}, {Morris}, {Mulone}, {Munoz},
  {Muraveva}, {Murphy}, {Musella}, {Noval}, {Ord{\'e}novic}, {Orr{\`u}},
  {Osinde}, {Pagani}, {Pagano}, {Palaversa}, {Palicio}, {Panahi}, {Pawlak},
  {Pe{\~n}alosa Esteller}, {Penttil{\"a}}, {Piersimoni}, {Pineau}, {Plachy},
  {Plum}, {Poggio}, {Poretti}, {Poujoulet}, {Pr{\v{s}}a}, {Pulone}, {Racero},
  {Ragaini}, {Rainer}, {Raiteri}, {Rambaux}, {Ramos}, {Ramos-Lerate}, {Re
  Fiorentin}, {Regibo}, {Reyl{\'e}}, {Ripepi}, {Riva}, {Rixon}, {Robichon},
  {Robin}, {Roelens}, {Rohrbasser}, {Romero-G{\'o}mez}, {Rowell}, {Royer},
  {Rybicki}, {Sadowski}, {Sagrist{\`a} Sell{\'e}s}, {Sahlmann}, {Salgado},
  {Salguero}, {Samaras}, {Sanchez Gimenez}, {Sanna}, {Santove{\~n}a},
  {Sarasso}, {Schultheis}, {Sciacca}, {Segol}, {Segovia}, {S{\'e}gransan},
  {Semeux}, {Shahaf}, {Siddiqui}, {Siebert}, {Siltala}, {Slezak}, {Smart},
  {Solano}, {Solitro}, {Souami}, {Souchay}, {Spagna}, {Spoto}, {Steele},
  {Steidelm{\"u}ller}, {Stephenson}, {S{\"u}veges}, {Szabados}, {Szegedi-Elek},
  {Taris}, {Tauran}, {Taylor}, {Teixeira}, {Thuillot}, {Tonello}, {Torra},
  {Torra}, {Turon}, {Unger}, {Vaillant}, {van Dillen}, {Vanel}, {Vecchiato},
  {Viala}, {Vicente}, {Voutsinas}, {Weiler}, {Wevers}, {Wyrzykowski}, {Yoldas},
  {Yvard}, {Zhao}, {Zorec}, {Zucker}, {Zurbach}, \&
  {Zwitter}}]{2021AandA...649A...1G}
{Gaia Collaboration}, {Brown}, A.~G.~A., {Vallenari}, A., {et~al.} 2021,
  \bibinfo{title}{{Gaia Early Data Release 3. Summary of the contents and
  survey properties},} \aap, 649, A1, \dodoi{10.1051/0004-6361/202039657}

\bibitem[{ {Gaia Collaboration} {et~al.}(2022){Gaia Collaboration},
  {Vallenari}, {Brown}, {Prusti}, {de Bruijne}, {Arenou}, {Babusiaux},
  {Biermann}, {Creevey}, {Ducourant}, \& et~al.}]{GaiaCollaboration2022}
{Gaia Collaboration}, {Vallenari}, A., {Brown}, A.~G.~A., {et~al.} 2022,
  \bibinfo{title}{{Gaia Data Release 3: Summary of the content and survey
  properties},} arXiv e-prints, arXiv:2208.00211,
  \dodoi{10.48550/arXiv.2208.00211}

\bibitem[{J.~H. Girard {et~al.}(2024)Girard, Gennaro, Rest, Leisenring,
  Hilbert, Canipe, Boyer, Golimowski, Sohn, Sahoo, Sunnquist, Brooks,
  Koekemoer, Kozhurina-Platais, Pierel, Carter, Stansberry, Perrin, Pueyo,
  Balmer, Kammerer, Lawson, Bogat, Wang, \& Rieke}]{Girard2024}
Girard, J.~H., Gennaro, M., Rest, A., {et~al.} 2024, in Space Telescopes and
  Instrumentation 2024: Optical, Infrared, and Millimeter Wave, ed. L.~E.
  Coyle, S.~Matsuura, \& M.~D. Perrin, Vol. 13092, International Society for
  Optics and Photonics (SPIE), 1309251, \dodoi{10.1117/12.3020243}

\bibitem[{J. {Golomb} {et~al.}(2021){Golomb}, {Rocha}, {Meshkat}, {Bottom},
  {Mawet}, {Mennesson}, {Vasisht}, \& {Wang}}]{Golomb2021}
{Golomb}, J., {Rocha}, G., {Meshkat}, T., {et~al.} 2021,
  \bibinfo{title}{{PlanetEvidence: Planet or Noise?},} \aj, 162, 304,
  \dodoi{10.3847/1538-3881/ac174e}

\bibitem[{G. {Gonzalez} {et~al.}(1999){Gonzalez}, {Wallerstein}, \&
  {Saar}}]{1999ApJ...511L.111G}
{Gonzalez}, G., {Wallerstein}, G., \& {Saar}, S.~H. 1999,
  \bibinfo{title}{{Parent Stars of Extrasolar Planets. IV. 14 Herculis, HD
  187123, and HD 210277},} \apjl, 511, L111, \dodoi{10.1086/311847}

\bibitem[{T.~D. {Groff} {et~al.}(2021){Groff}, {Zimmerman}, {Subedi}, {Rizzo},
  {Titus}, {Lyons}, {Bell}, {Gaylin}, {Gao}, {Pasquale}, {Nicolaeff}, {Tamura},
  \& {Shi}}]{2021SPIE11443E..3DG}
{Groff}, T.~D., {Zimmerman}, N.~T., {Subedi}, H.~B., {et~al.} 2021, in Society
  of Photo-Optical Instrumentation Engineers (SPIE) Conference Series, Vol.
  11443, Society of Photo-Optical Instrumentation Engineers (SPIE) Conference
  Series, ed. M.~{Lystrup} \& M.~D. {Perrin}, 114433D,
  \dodoi{10.1117/12.2562925}

\bibitem[{R. {Hanel} {et~al.}(1981){Hanel}, {Conrath}, {Herath}, {Kunde}, \&
  {Pirraglia}}]{1981JGR....86.8705H}
{Hanel}, R., {Conrath}, B., {Herath}, L., {Kunde}, V., \& {Pirraglia}, J. 1981,
  \bibinfo{title}{{Albedo, internal heat, and energy balance of Jupiter -
  Preliminary results of the Voyager infrared investigation},} \jgr, 86, 8705,
  \dodoi{10.1029/JA086iA10p08705}

\bibitem[{S. {Hinkley} {et~al.}(2023){Hinkley}, {Lacour}, {Marleau},
  {Lagrange}, {Wang}, {Kammerer}, {Cumming}, {Nowak}, {Rodet}, {Stolker},
  {Balmer}, {Ray}, {Bonnefoy}, {Molli{\`e}re}, {Lazzoni}, {Kennedy},
  {Mordasini}, {Abuter}, {Aigrain}, {Amorim}, {Asensio-Torres}, {Babusiaux},
  {Benisty}, {Berger}, {Beust}, {Blunt}, {Boccaletti}, {Bohn}, {Bonnet},
  {Bourdarot}, {Brandner}, {Cantalloube}, {Caselli}, {Charnay}, {Chauvin},
  {Chomez}, {Choquet}, {Christiaens}, {Cl{\'e}net}, {Coud{\'e} du Foresto},
  {Cridland}, {Delorme}, {Dembet}, {Drescher}, {Duvert}, {Eckart},
  {Eisenhauer}, {Feuchtgruber}, {Galland}, {Garcia}, {Garcia Lopez}, {Gardner},
  {Gendron}, {Genzel}, {Gillessen}, {Girard}, {Grandjean}, {Haubois},
  {Hei{\ss}el}, {Henning}, {Hippler}, {Horrobin}, {Houll{\'e}}, {Hubert},
  {Jocou}, {Keppler}, {Kervella}, {Kreidberg}, {Lapeyr{\`e}re}, {Le Bouquin},
  {L{\'e}na}, {Lutz}, {Maire}, {Mang}, {M{\'e}rand}, {Meunier}, {Monnier},
  {Mouillet}, {Nasedkin}, {Ott}, {Otten}, {Paladini}, {Paumard}, {Perraut},
  {Perrin}, {Philipot}, {Pfuhl}, {Pourr{\'e}}, {Pueyo}, {Rameau}, {Rickman},
  {Rubini}, {Rustamkulov}, {Samland}, {Shangguan}, {Shimizu}, {Sing},
  {Straubmeier}, {Sturm}, {Tacconi}, {van Dishoeck}, {Vigan}, {Vincent},
  {Ward-Duong}, {Widmann}, {Wieprecht}, {Wiezorrek}, {Woillez}, {Yazici},
  {Young}, \& {Zicher}}]{Hinkley2023}
{Hinkley}, S., {Lacour}, S., {Marleau}, G.~D., {et~al.} 2023,
  \bibinfo{title}{{Direct discovery of the inner exoplanet in the HD 206893
  system. Evidence for deuterium burning in a planetary-mass companion},} \aap,
  671, L5, \dodoi{10.1051/0004-6361/202244727}

\bibitem[{E. {H{\o}g} {et~al.}(2000){H{\o}g}, {Fabricius}, {Makarov}, {Urban},
  {Corbin}, {Wycoff}, {Bastian}, {Schwekendiek}, \&
  {Wicenec}}]{2000A&A...355L..27H}
{H{\o}g}, E., {Fabricius}, C., {Makarov}, V.~V., {et~al.} 2000,
  \bibinfo{title}{{The Tycho-2 catalogue of the 2.5 million brightest stars},}
  \aap, 355, L27

\bibitem[{J. {Kammerer} {et~al.}(2022){Kammerer}, {Girard}, {Carter}, {Perrin},
  {Cooper}, {Thatte}, {Vandal}, {Leisenring}, {Wang}, {Balmer},
  {Sivaramakrishnan}, {Pueyo}, {Ward-Duong}, {Sunnquist}, \& {Adams
  Redai}}]{Kammerer2022}
{Kammerer}, J., {Girard}, J., {Carter}, A.~L., {et~al.} 2022, in Society of
  Photo-Optical Instrumentation Engineers (SPIE) Conference Series, Vol. 12180,
  Space Telescopes and Instrumentation 2022: Optical, Infrared, and Millimeter
  Wave, ed. L.~E. {Coyle}, S.~{Matsuura}, \& M.~D. {Perrin}, 121803N,
  \dodoi{10.1117/12.2628865}

\bibitem[{J. {Kammerer} {et~al.}(2024){Kammerer}, {Lawson}, {Perrin},
  {Rebollido}, {Stark}, {Stolker}, {Girard}, {Pueyo}, {Balmer}, {Worthen},
  {Chen}, {van der Marel}, {Lewis}, {Ward-Duong}, {Valenti}, {Clampin}, \&
  {Mountain}}]{Kammerer2024}
{Kammerer}, J., {Lawson}, K., {Perrin}, M.~D., {et~al.} 2024,
  \bibinfo{title}{{JWST-TST High Contrast: JWST/NIRCam observations of the
  young giant planet $\beta$ Pic b},} arXiv e-prints, arXiv:2405.18422,
  \dodoi{10.48550/arXiv.2405.18422}

\bibitem[{J.~D. {Kirkpatrick} {et~al.}(2024){Kirkpatrick}, {Marocco}, {Gelino},
  {Raghu}, {Faherty}, {Bardalez Gagliuffi}, {Schurr}, {Apps}, {Schneider},
  {Meisner}, {Kuchner}, {Caselden}, {Smart}, {Casewell}, {Raddi}, {Kesseli},
  {Stevnbak Andersen}, {Antonini}, {Beaulieu}, {Bickle}, {Bilsing}, {Chieng},
  {Colin}, {Deen}, {Dereveanco}, {Doll}, {Durantini Luca}, {Frazer}, {Gantier},
  {Gramaize}, {Grant}, {Hamlet}, {Higashimura}, {Hyogo}, {Ja{\l}owiczor},
  {Jonkeren}, {Kabatnik}, {Kiwy}, {Martin}, {Michaels}, {Pendrill}, {Pessanha
  Machado}, {Pumphrey}, {Rothermich}, {Russwurm}, {Sainio}, {Sanchez},
  {Sapelkin-Tambling}, {Sch{\"u}mann}, {Selg-Mann}, {Singh}, {Stenner}, {Sun},
  {Tanner}, {Th{\'e}venot}, {Ventura}, {Voloshin}, {Walla}, {W{\k{e}}dracki},
  {Adorno}, {Aganze}, {Allers}, {Brooks}, {Burgasser}, {Calamari}, {Connor},
  {Costa}, {Eisenhardt}, {Gagn{\'e}}, {Gerasimov}, {Gonzales}, {Hsu}, {Kiman},
  {Li}, {Low}, {Mamajek}, {Pantoja}, {Popinchalk}, {Rees}, {Stern},
  {Su{\'a}rez}, {Theissen}, {Tsai}, {Vos}, {Zurek}, \& {The Backyard Worlds:
  Planet 9 Collaboration}}]{2024ApJS..271...55K}
{Kirkpatrick}, J.~D., {Marocco}, F., {Gelino}, C.~R., {et~al.} 2024,
  \bibinfo{title}{{The Initial Mass Function Based on the Full-sky 20 pc Census
  of {\ensuremath{\sim}}3600 Stars and Brown Dwarfs},} \apjs, 271, 55,
  \dodoi{10.3847/1538-4365/ad24e2}

\bibitem[{Y. {Kozai}(1962){Kozai}}]{1962AJ.....67R.579K}
{Kozai}, Y. 1962, \bibinfo{title}{{Secular Perturbations of Asteroids with High
  Inclination and Eccentricity.},} \aj, 67, 579, \dodoi{10.1086/108876}

\bibitem[{J.~E. {Krist} {et~al.}(2010){Krist}, {Balasubramanian}, {Muller},
  {Shaklan}, {Kelly}, {Wilson}, {Beichman}, {Serabyn}, {Mao}, {Echternach},
  {Trauger}, \& {Liewer}}]{Krist2010}
{Krist}, J.~E., {Balasubramanian}, K., {Muller}, R.~E., {et~al.} 2010, in
  Society of Photo-Optical Instrumentation Engineers (SPIE) Conference Series,
  Vol. 7731, Space Telescopes and Instrumentation 2010: Optical, Infrared, and
  Millimeter Wave, ed. J.~{Oschmann}, Jacobus~M., M.~C. {Clampin}, \& H.~A.
  {MacEwen}, 77313J, \dodoi{10.1117/12.856488}

\bibitem[{S. {Lacour} {et~al.}(2021){Lacour}, {Wang}, {Rodet}, {Nowak},
  {Shangguan}, {Beust}, {Lagrange}, {Abuter}, {Amorim}, {Asensio-Torres},
  {Benisty}, {Berger}, {Blunt}, {Boccaletti}, {Bohn}, {Bolzer}, {Bonnefoy},
  {Bonnet}, {Bourdarot}, {Brandner}, {Cantalloube}, {Caselli}, {Charnay},
  {Chauvin}, {Choquet}, {Christiaens}, {Cl{\'e}net}, {Coud{\'e} Du Foresto},
  {Cridland}, {Dembet}, {Dexter}, {de Zeeuw}, {Drescher}, {Duvert}, {Eckart},
  {Eisenhauer}, {Gao}, {Garcia}, {Garcia Lopez}, {Gendron}, {Genzel},
  {Gillessen}, {Girard}, {Haubois}, {Hei{\ss}el}, {Henning}, {Hinkley},
  {Hippler}, {Horrobin}, {Houll{\'e}}, {Hubert}, {Jocou}, {Kammerer},
  {Keppler}, {Kervella}, {Kreidberg}, {Lapeyr{\`e}re}, {Le Bouquin},
  {L{\'e}na}, {Lutz}, {Maire}, {M{\'e}rand}, {Molli{\`e}re}, {Monnier},
  {Mouillet}, {Nasedkin}, {Ott}, {Otten}, {Paladini}, {Paumard}, {Perraut},
  {Perrin}, {Pfuhl}, {Rickman}, {Pueyo}, {Rameau}, {Rousset}, {Rustamkulov},
  {Samland}, {Shimizu}, {Sing}, {Stadler}, {Stolker}, {Straub}, {Straubmeier},
  {Sturm}, {Tacconi}, {van Dishoeck}, {Vigan}, {Vincent}, {von Fellenberg},
  {Ward-Duong}, {Widmann}, {Wieprecht}, {Wiezorrek}, {Woillez}, {Yazici},
  {Young}, \& {Gravity Collaboration}}]{Lacour2021}
{Lacour}, S., {Wang}, J.~J., {Rodet}, L., {et~al.} 2021, \bibinfo{title}{{The
  mass of {\ensuremath{\beta}} Pictoris c from {\ensuremath{\beta}} Pictoris b
  orbital motion},} \aap, 654, L2, \dodoi{10.1051/0004-6361/202141889}

\bibitem[{B. {Lacy} \& A. {Burrows}(2023){Lacy} \&
  {Burrows}}]{2023ApJ...950....8L}
{Lacy}, B., \& {Burrows}, A. 2023, \bibinfo{title}{{Self-consistent Models of Y
  Dwarf Atmospheres with Water Clouds and Disequilibrium Chemistry},} \apj,
  950, 8, \dodoi{10.3847/1538-4357/acc8cb}

\bibitem[{A.~M. {Lagrange} {et~al.}(2010){Lagrange}, {Bonnefoy}, {Chauvin},
  {Apai}, {Ehrenreich}, {Boccaletti}, {Gratadour}, {Rouan}, {Mouillet},
  {Lacour}, \& {Kasper}}]{2010Sci...329...57L}
{Lagrange}, A.~M., {Bonnefoy}, M., {Chauvin}, G., {et~al.} 2010,
  \bibinfo{title}{{A Giant Planet Imaged in the Disk of the Young Star
  {\ensuremath{\beta}} Pictoris},} Science, 329, 57,
  \dodoi{10.1126/science.1187187}

\bibitem[{A.~M. {Lagrange} {et~al.}(2025{\natexlab{a}}){Lagrange}, {Wilkinson},
  {M{\^a}lin}, {Boccaletti}, {Perrot}, {Matr{\`a}}, {Combes}, {Rouan}, {Beust},
  {Chomez}, {Charnay}, {Mazevet}, {Flasseur}, {Olofsson}, {Bayo}, {Kral},
  {Chauvin}, {Thebault}, {Rubini}, {Milli}, {Kiefer}, {Carter}, {Crotts},
  {Radcliffe}, {Mazoyer}, {Bodrito}, {Stasevic}, {Delorme}, \&
  {Langlois}}]{2025arXiv250215081L}
{Lagrange}, A.~M., {Wilkinson}, C., {M{\^a}lin}, M., {et~al.}
  2025{\natexlab{a}}, \bibinfo{title}{{Evidence for a sub-jovian planet in the
  young TWA7 disk},} arXiv e-prints, arXiv:2502.15081,
  \dodoi{10.48550/arXiv.2502.15081}

\bibitem[{A.~M. {Lagrange} {et~al.}(2025{\natexlab{b}}){Lagrange}, {Wilkinson},
  {M{\^a}lin}, {Boccaletti}, {Perrot}, {Matr{\`a}}, {Combes}, {Rouan}, {Beust},
  {Chomez}, {Charnay}, {Mazevet}, {Flasseur}, {Olofsson}, {Bayo}, {Kral},
  {Chauvin}, {Thebault}, {Rubini}, {Milli}, {Kiefer}, {Carter}, {Crotts},
  {Radcliffe}, {Mazoyer}, {Bodrito}, {Stasevic}, {Delorme}, \&
  {Langlois}}]{Lagrange2025}
{Lagrange}, A.~M., {Wilkinson}, C., {M{\^a}lin}, M., {et~al.}
  2025{\natexlab{b}}, \bibinfo{title}{{Evidence for a sub-jovian planet in the
  young TWA7 disk},} arXiv e-prints, arXiv:2502.15081,
  \dodoi{10.48550/arXiv.2502.15081}

\bibitem[{J. {Leisenring} {et~al.}(2022){Leisenring}, {Schlawin}, \&
  {Fraine}}]{pynrc}
{Leisenring}, J., {Schlawin}, E., \& {Fraine}, J. 2022,
  \bibinfo{title}{{JarronL/pynrc: Release v1.0.4},}, v1.0.4, Zenodo Zenodo,
  \dodoi{10.5281/zenodo.5829553}

\bibitem[{M.~L. {Lidov}(1962){Lidov}}]{Lidov1962}
{Lidov}, M.~L. 1962, \bibinfo{title}{{The evolution of orbits of artificial
  satellites of planets under the action of gravitational perturbations of
  external bodies},} \planss, 9, 719, \dodoi{10.1016/0032-0633(62)90129-0}

\bibitem[{L. Lindegren(2018)Lindegren}]{Lindegren2018}
Lindegren, L. 2018, \bibinfo{title}{Re-Normalising the Astrometric Chi-Square
  in {{Gaia DR2}},}, Tech. Rep. GAIA-C3-TN-LU-LL-124-01

\bibitem[{T. {Lu} {et~al.}(2025){Lu}, {An}, {Li}, {Millholland}, {Rice},
  {Brandt}, \& {Brandt}}]{Lu2025}
{Lu}, T., {An}, Q., {Li}, G., {et~al.} 2025,
  \bibinfo{title}{{Planet{\textendash}Planet Scattering and Von
  Zeipel{\textendash}Lidov{\textendash}Kozai Migration{\textemdash}The
  Dynamical History of HAT-P-11},} \apj, 979, 218,
  \dodoi{10.3847/1538-4357/ad9b79}

\bibitem[{R.~E. {Luck} \& U. {Heiter}(2006){Luck} \&
  {Heiter}}]{2006AJ....131.3069L}
{Luck}, R.~E., \& {Heiter}, U. 2006, \bibinfo{title}{{Dwarfs in the Local
  Region},} \aj, 131, 3069, \dodoi{10.1086/504080}

\bibitem[{K.~L. {Luhman} \& R. {Jayawardhana}(2002){Luhman} \&
  {Jayawardhana}}]{Luhman2002}
{Luhman}, K.~L., \& {Jayawardhana}, R. 2002, \bibinfo{title}{{An Adaptive
  Optics Search for Companions to Stars with Planets},} \apj, 566, 1132,
  \dodoi{10.1086/338338}

\bibitem[{K.~E. Mandt {et~al.}(2024)Mandt, Lustig-Yaeger, Luspay-Kuti, Wurz,
  Bodewits, Fuselier, Mousis, Petrinec, \&
  Trattner}]{doi:10.1126/sciadv.adp2191}
Mandt, K.~E., Lustig-Yaeger, J., Luspay-Kuti, A., {et~al.} 2024,
  \bibinfo{title}{A nearly terrestrial D/H for comet
  67P/Churyumov-Gerasimenko,} Science Advances, 10, eadp2191,
  \dodoi{10.1126/sciadv.adp2191}

\bibitem[{M.~S. {Marley} {et~al.}(2021{\natexlab{a}}){Marley}, {Saumon},
  {Visscher}, {Lupu}, {Freedman}, {Morley}, {Fortney}, {Seay}, {Smith}, {Teal},
  \& {Wang}}]{Marley2021}
{Marley}, M.~S., {Saumon}, D., {Visscher}, C., {et~al.} 2021{\natexlab{a}},
  \bibinfo{title}{{The Sonora Brown Dwarf Atmosphere and Evolution Models. I.
  Model Description and Application to Cloudless Atmospheres in Rainout
  Chemical Equilibrium},} \apj, 920, 85, \dodoi{10.3847/1538-4357/ac141d}

\bibitem[{M.~S. {Marley} {et~al.}(2021{\natexlab{b}}){Marley}, {Saumon},
  {Visscher}, {Lupu}, {Freedman}, {Morley}, {Fortney}, {Seay}, {Smith}, {Teal},
  \& {Wang}}]{2021ApJ...920...85M}
{Marley}, M.~S., {Saumon}, D., {Visscher}, C., {et~al.} 2021{\natexlab{b}},
  \bibinfo{title}{{The Sonora Brown Dwarf Atmosphere and Evolution Models. I.
  Model Description and Application to Cloudless Atmospheres in Rainout
  Chemical Equilibrium},} \apj, 920, 85, \dodoi{10.3847/1538-4357/ac141d}

\bibitem[{C. {Marois} {et~al.}(2008){Marois}, {Macintosh}, {Barman},
  {Zuckerman}, {Song}, {Patience}, {Lafreni{\`e}re}, \&
  {Doyon}}]{2008Sci...322.1348M}
{Marois}, C., {Macintosh}, B., {Barman}, T., {et~al.} 2008,
  \bibinfo{title}{{Direct Imaging of Multiple Planets Orbiting the Star HR
  8799},} Science, 322, 1348, \dodoi{10.1126/science.1166585}

\bibitem[{E.~C. {Matthews} {et~al.}(2024{\natexlab{a}}){Matthews}, {Carter},
  {Pathak}, {Morley}, {Phillips}, {P.~M.}, {Feng}, {Bonse}, {Boogaard}, {Burt},
  {Crossfield}, {Douglas}, {Henning}, {Hom}, {Ko}, {Kasper}, {Lagrange}, {Petit
  dit de la Roche}, \& {Philipot}}]{2024Natur.633..789M}
{Matthews}, E.~C., {Carter}, A.~L., {Pathak}, P., {et~al.} 2024{\natexlab{a}},
  \bibinfo{title}{{A temperate super-Jupiter imaged with JWST in the
  mid-infrared},} \nat, 633, 789, \dodoi{10.1038/s41586-024-07837-8}

\bibitem[{E.~C. {Matthews} {et~al.}(2024{\natexlab{b}}){Matthews}, {Carter},
  {Pathak}, {Morley}, {Phillips}, {P.~M.}, {Feng}, {Bonse}, {Boogaard}, {Burt},
  {Crossfield}, {Douglas}, {Henning}, {Hom}, {Ko}, {Kasper}, {Lagrange}, {Petit
  dit de la Roche}, \& {Philipot}}]{Matthews2024}
{Matthews}, E.~C., {Carter}, A.~L., {Pathak}, P., {et~al.} 2024{\natexlab{b}},
  \bibinfo{title}{{A temperate super-Jupiter imaged with JWST in the
  mid-infrared},} \nat, 633, 789, \dodoi{10.1038/s41586-024-07837-8}

\bibitem[{D. {Mawet} {et~al.}(2014){Mawet}, {Milli}, {Wahhaj}, {Pelat},
  {Absil}, {Delacroix}, {Boccaletti}, {Kasper}, {Kenworthy}, {Marois},
  {Mennesson}, \& {Pueyo}}]{Mawet2014}
{Mawet}, D., {Milli}, J., {Wahhaj}, Z., {et~al.} 2014,
  \bibinfo{title}{{Fundamental Limitations of High Contrast Imaging Set by
  Small Sample Statistics},} \apj, 792, 97, \dodoi{10.1088/0004-637X/792/2/97}

\bibitem[{B.~E. {McArthur} {et~al.}(2010){McArthur}, {Benedict}, {Barnes},
  {Martioli}, {Korzennik}, {Nelan}, \& {Butler}}]{2010ApJ...715.1203M}
{McArthur}, B.~E., {Benedict}, G.~F., {Barnes}, R., {et~al.} 2010,
  \bibinfo{title}{{New Observational Constraints on the {\ensuremath{\upsilon}}
  Andromedae System with Data from the Hubble Space Telescope and Hobby-Eberly
  Telescope},} \apj, 715, 1203, \dodoi{10.1088/0004-637X/715/2/1203}

\bibitem[{E. {Merlin} {et~al.}(2024){Merlin}, {Santini}, {Paris}, {Castellano},
  {Fontana}, {Treu}, {Finkelstein}, {Dunlop}, {Arrabal Haro}, {Bagley},
  {Boyett}, {Calabr{\`o}}, {Correnti}, {Davis}, {Dickinson}, {Donnan},
  {Ferguson}, {Fortuni}, {Giavalisco}, {Glazebrook}, {Grazian}, {Grogin},
  {Hathi}, {Hirschmann}, {Kartaltepe}, {Kewley}, {Kirkpatrick}, {Kocevski},
  {Koekemoer}, {Leung}, {Lotz}, {Lucas}, {Magee}, {Marchesini}, {Mascia},
  {McLeod}, {McLure}, {Nanayakkara}, {Napolitano}, {Nonino}, {Papovich},
  {Pentericci}, {P{\'e}rez-Gonz{\'a}lez}, {Pirzkal}, {Ravindranath},
  {Roberts-Borsani}, {Somerville}, {Trenti}, {Trump}, {Vulcani}, {Wang},
  {Watson}, {Wilkins}, {Yang}, \& {Yung}}]{2024A&A...691A.240M}
{Merlin}, E., {Santini}, P., {Paris}, D., {et~al.} 2024,
  \bibinfo{title}{{ASTRODEEP-JWST: NIRCam-HST multi-band photometry and
  redshifts for half a million sources in six extragalactic deep fields},}
  \aap, 691, A240, \dodoi{10.1051/0004-6361/202451409}

\bibitem[{B.~E. {Miles} {et~al.}(2020){Miles}, {Skemer}, {Morley}, {Marley},
  {Fortney}, {Allers}, {Faherty}, {Geballe}, {Visscher}, {Schneider}, {Lupu},
  {Freedman}, \& {Bjoraker}}]{2020AJ....160...63M}
{Miles}, B.~E., {Skemer}, A. J.~I., {Morley}, C.~V., {et~al.} 2020,
  \bibinfo{title}{{Observations of Disequilibrium CO Chemistry in the Coldest
  Brown Dwarfs},} \aj, 160, 63, \dodoi{10.3847/1538-3881/ab9114}

\bibitem[{S.~M. {Mills} \& D.~C. {Fabrycky}(2017){Mills} \&
  {Fabrycky}}]{2017AJ....153...45M}
{Mills}, S.~M., \& {Fabrycky}, D.~C. 2017, \bibinfo{title}{{Kepler-108: A
  Mutually Inclined Giant Planet System},} \aj, 153, 45,
  \dodoi{10.3847/1538-3881/153/1/45}

\bibitem[{C.~V. {Morley} {et~al.}(2015){Morley}, {Fortney}, {Marley}, {Zahnle},
  {Line}, {Kempton}, {Lewis}, \& {Cahoy}}]{Morley2015}
{Morley}, C.~V., {Fortney}, J.~J., {Marley}, M.~S., {et~al.} 2015,
  \bibinfo{title}{{Thermal Emission and Reflected Light Spectra of Super Earths
  with Flat Transmission Spectra},} \apj, 815, 110,
  \dodoi{10.1088/0004-637X/815/2/110}

\bibitem[{C.~V. {Morley} {et~al.}(2024){Morley}, {Mukherjee}, {Marley},
  {Fortney}, {Visscher}, {Lupu}, {Gharib-Nezhad}, {Thorngren}, {Freedman}, \&
  {Batalha 7}}]{Morley2024}
{Morley}, C.~V., {Mukherjee}, S., {Marley}, M.~S., {et~al.} 2024,
  \bibinfo{title}{{The Sonora Substellar Atmosphere Models. III. Diamondback:
  Atmospheric Properties, Spectra, and Evolution for Warm Cloudy Substellar
  Objects},} arXiv e-prints, arXiv:2402.00758,
  \dodoi{10.48550/arXiv.2402.00758}

\bibitem[{S. {Mukherjee} {et~al.}(2023){Mukherjee}, {Batalha}, {Fortney}, \&
  {Marley}}]{Mukherjee2023}
{Mukherjee}, S., {Batalha}, N.~E., {Fortney}, J.~J., \& {Marley}, M.~S. 2023,
  \bibinfo{title}{{PICASO 3.0: A One-dimensional Climate Model for Giant
  Planets and Brown Dwarfs},} \apj, 942, 71, \dodoi{10.3847/1538-4357/ac9f48}

\bibitem[{S. {Mukherjee} {et~al.}(2024{\natexlab{a}}){Mukherjee}, {Fortney},
  {Wogan}, {Sing}, \& {Ohno}}]{Mukherjee2024Photochem}
{Mukherjee}, S., {Fortney}, J.~J., {Wogan}, N.~F., {Sing}, D.~K., \& {Ohno}, K.
  2024{\natexlab{a}}, \bibinfo{title}{{Effects of Planetary Parameters on
  Disequilibrium Chemistry in Irradiated Planetary Atmospheres: From Gas Giants
  to Sub-Neptunes},} arXiv e-prints, arXiv:2410.17169,
  \dodoi{10.48550/arXiv.2410.17169}

\bibitem[{S. {Mukherjee} {et~al.}(2024{\natexlab{b}}){Mukherjee}, {Fortney},
  {Morley}, {Batalha}, {Marley}, {Karalidi}, {Visscher}, {Lupu}, {Freedman}, \&
  {Gharib-Nezhad}}]{2024ApJ...963...73M}
{Mukherjee}, S., {Fortney}, J.~J., {Morley}, C.~V., {et~al.}
  2024{\natexlab{b}}, \bibinfo{title}{{The Sonora Substellar Atmosphere Models.
  IV. Elf Owl: Atmospheric Mixing and Chemical Disequilibrium with Varying
  Metallicity and C/O Ratios},} \apj, 963, 73, \dodoi{10.3847/1538-4357/ad18c2}

\bibitem[{S. {Mukherjee} {et~al.}(2024{\natexlab{c}}){Mukherjee}, {Fortney},
  {Morley}, {Batalha}, {Marley}, {Karalidi}, {Visscher}, {Lupu}, {Freedman}, \&
  {Gharib-Nezhad}}]{Mukherjee2024EO}
{Mukherjee}, S., {Fortney}, J.~J., {Morley}, C.~V., {et~al.}
  2024{\natexlab{c}}, \bibinfo{title}{{The Sonora Substellar Atmosphere Models.
  IV. Elf Owl: Atmospheric Mixing and Chemical Disequilibrium with Varying
  Metallicity and C/O Ratios},} \apj, 963, 73, \dodoi{10.3847/1538-4357/ad18c2}

\bibitem[{D. {Naef} {et~al.}(2004){Naef}, {Mayor}, {Beuzit}, {Perrier},
  {Queloz}, {Sivan}, \& {Udry}}]{Naef2004}
{Naef}, D., {Mayor}, M., {Beuzit}, J.~L., {et~al.} 2004, \bibinfo{title}{{The
  ELODIE survey for northern extra-solar planets. III. Three planetary
  candidates detected with ELODIE},} \aap, 414, 351,
  \dodoi{10.1051/0004-6361:20034091}

\bibitem[{S. {Naoz}(2016){Naoz}}]{2016ARA&A..54..441N}
{Naoz}, S. 2016, \bibinfo{title}{{The Eccentric Kozai-Lidov Effect and Its
  Applications},} \araa, 54, 441, \dodoi{10.1146/annurev-astro-081915-023315}

\bibitem[{S. {Naoz} {et~al.}(2011){Naoz}, {Farr}, {Lithwick}, {Rasio}, \&
  {Teyssandier}}]{2011Natur.473..187N}
{Naoz}, S., {Farr}, W.~M., {Lithwick}, Y., {Rasio}, F.~A., \& {Teyssandier}, J.
  2011, \bibinfo{title}{{Hot Jupiters from secular planet-planet
  interactions},} \nat, 473, 187, \dodoi{10.1038/nature10076}

\bibitem[{S. {Naoz} {et~al.}(2013){Naoz}, {Farr}, {Lithwick}, {Rasio}, \&
  {Teyssandier}}]{naoz2013}
{Naoz}, S., {Farr}, W.~M., {Lithwick}, Y., {Rasio}, F.~A., \& {Teyssandier}, J.
  2013, \bibinfo{title}{{Secular dynamics in hierarchical three-body systems},}
  \mnras, 431, 2155, \dodoi{10.1093/mnras/stt302}

\bibitem[{D. {Nesvorn{\'y}}(2011){Nesvorn{\'y}}}]{2011ApJ...742L..22N}
{Nesvorn{\'y}}, D. 2011, \bibinfo{title}{{Young Solar System's Fifth Giant
  Planet?},} \apjl, 742, L22, \dodoi{10.1088/2041-8205/742/2/L22}

\bibitem[{J. {Patience} {et~al.}(2002){Patience}, {White}, {Ghez}, {McCabe},
  {McLean}, {Larkin}, {Prato}, {Kim}, {Lloyd}, {Liu}, {Graham}, {Macintosh},
  {Gavel}, {Max}, {Bauman}, {Olivier}, {Wizinowich}, \& {Acton}}]{Patience2002}
{Patience}, J., {White}, R.~J., {Ghez}, A.~M., {et~al.} 2002,
  \bibinfo{title}{{Stellar Companions to Stars with Planets},} \apj, 581, 654,
  \dodoi{10.1086/342982}

\bibitem[{L. {Pueyo}(2016){Pueyo}}]{Pueyo2016}
{Pueyo}, L. 2016, \bibinfo{title}{{Detection and Characterization of Exoplanets
  using Projections on Karhunen Loeve Eigenimages: Forward Modeling},} \apj,
  824, 117, \dodoi{10.3847/0004-637X/824/2/117}

\bibitem[{S.~N. {Raymond}(2024){Raymond}}]{2024arXiv240414982R}
{Raymond}, S.~N. 2024, \bibinfo{title}{{The Solar System: structural overview,
  origins and evolution},} arXiv e-prints, arXiv:2404.14982,
  \dodoi{10.48550/arXiv.2404.14982}

\bibitem[{M.~J. {Rieke} {et~al.}(2023){Rieke}, {Robertson}, {Tacchella},
  {Hainline}, {Johnson}, {Hausen}, {Ji}, {Willmer}, {Eisenstein}, {Pusk{\'a}s},
  {Alberts}, {Arribas}, {Baker}, {Baum}, {Bhatawdekar}, {Bonaventura},
  {Boyett}, {Bunker}, {Cameron}, {Carniani}, {Charlot}, {Chevallard}, {Chen},
  {Curti}, {Curtis-Lake}, {Danhaive}, {DeCoursey}, {Dressler}, {Egami},
  {Endsley}, {Helton}, {Hviding}, {Kumari}, {Looser}, {Lyu}, {Maiolino},
  {Maseda}, {Nelson}, {Rieke}, {Rix}, {Sandles}, {Saxena}, {Sharpe}, {Shivaei},
  {Skarbinski}, {Smit}, {Stark}, {Stone}, {Suess}, {Sun}, {Topping},
  {{\"U}bler}, {Villanueva}, {Wallace}, {Williams}, {Willott}, {Whitler},
  {Witstok}, \& {Woodrum}}]{Rieke2023}
{Rieke}, M.~J., {Robertson}, B., {Tacchella}, S., {et~al.} 2023,
  \bibinfo{title}{{JADES Initial Data Release for the Hubble Ultra Deep Field:
  Revealing the Faint Infrared Sky with Deep JWST NIRCam Imaging},} \apjs, 269,
  16, \dodoi{10.3847/1538-4365/acf44d}

\bibitem[{J. {Rigby} {et~al.}(2023){Rigby}, {Perrin}, {McElwain}, {Kimble},
  {Friedman}, {Lallo}, {Doyon}, {Feinberg}, {Ferruit}, {Glasse}, \&
  et~al.}]{Rigby2023}
{Rigby}, J., {Perrin}, M., {McElwain}, M., {et~al.} 2023, \bibinfo{title}{{The
  Science Performance of JWST as Characterized in Commissioning},} \pasp, 135,
  048001, \dodoi{10.1088/1538-3873/acb293}

\bibitem[{T.~J. {Rodigas} {et~al.}(2011){Rodigas}, {Males}, {Hinz}, {Mamajek},
  \& {Knox}}]{Rodigas2011}
{Rodigas}, T.~J., {Males}, J.~R., {Hinz}, P.~M., {Mamajek}, E.~E., \& {Knox},
  R.~P. 2011, \bibinfo{title}{{Direct Imaging Constraints on the Putative
  Exoplanet 14 Her C},} \apj, 732, 10, \dodoi{10.1088/0004-637X/732/1/10}

\bibitem[{L.~J. {Rosenthal} {et~al.}(2021){Rosenthal}, {Fulton}, {Hirsch},
  {Isaacson}, {Howard}, {Dedrick}, {Sherstyuk}, {Blunt}, {Petigura}, {Knutson},
  {Behmard}, {Chontos}, {Crepp}, {Crossfield}, {Dalba}, {Fischer}, {Henry},
  {Kane}, {Kosiarek}, {Marcy}, {Rubenzahl}, {Weiss}, \&
  {Wright}}]{Rosenthal2021}
{Rosenthal}, L.~J., {Fulton}, B.~J., {Hirsch}, L.~A., {et~al.} 2021,
  \bibinfo{title}{{The California Legacy Survey. I. A Catalog of 178 Planets
  from Precision Radial Velocity Monitoring of 719 Nearby Stars over Three
  Decades},} \apjs, 255, 8, \dodoi{10.3847/1538-4365/abe23c}

\bibitem[{M.~J. {Rowland} {et~al.}(2024){Rowland}, {Morley}, {Miles}, {Suarez},
  {Faherty}, {Skemer}, {Beiler}, {Line}, {Bjoraker}, {Fortney}, {Vos},
  {Alejandro Merchan}, {Marley}, {Burningham}, {Freedman}, {Gharib-Nezhad},
  {Batalha}, {Lupu}, {Visscher}, {Schneider}, {Geballe}, {Carter}, {Allers},
  {Mang}, {Apai}, {Limbach}, \& {Wilson}}]{2024ApJ...977L..49R}
{Rowland}, M.~J., {Morley}, C.~V., {Miles}, B.~E., {et~al.} 2024,
  \bibinfo{title}{{Protosolar D-to-H Abundance and One Part per Billion
  PH$_{3}$ in the Coldest Brown Dwarf},} \apjl, 977, L49,
  \dodoi{10.3847/2041-8213/ad9744}

\bibitem[{D. {Saumon} \& M.~S. {Marley}(2008){Saumon} \& {Marley}}]{Saumon2008}
{Saumon}, D., \& {Marley}, M.~S. 2008, \bibinfo{title}{{The Evolution of L and
  T Dwarfs in Color-Magnitude Diagrams},} \apj, 689, 1327,
  \dodoi{10.1086/592734}

\bibitem[{A.~J. {Skemer} {et~al.}(2016){Skemer}, {Morley}, {Allers}, {Geballe},
  {Marley}, {Fortney}, {Faherty}, {Bjoraker}, \& {Lupu}}]{skemer2016}
{Skemer}, A.~J., {Morley}, C.~V., {Allers}, K.~N., {et~al.} 2016,
  \bibinfo{title}{{The First Spectrum of the Coldest Brown Dwarf},} \apjl, 826,
  L17, \dodoi{10.3847/2041-8205/826/2/L17}

\bibitem[{M.~F. {Skrutskie} {et~al.}(2006){Skrutskie}, {Cutri}, {Stiening},
  {Weinberg}, {Schneider}, {Carpenter}, {Beichman}, {Capps}, {Chester},
  {Elias}, {Huchra}, {Liebert}, {Lonsdale}, {Monet}, {Price}, {Seitzer},
  {Jarrett}, {Kirkpatrick}, {Gizis}, {Howard}, {Evans}, {Fowler}, {Fullmer},
  {Hurt}, {Light}, {Kopan}, {Marsh}, {McCallon}, {Tam}, {Van Dyk}, \&
  {Wheelock}}]{skr06}
{Skrutskie}, M.~F., {Cutri}, R.~M., {Stiening}, R., {et~al.} 2006,
  \bibinfo{title}{{The Two Micron All Sky Survey (2MASS)},} \aj, 131, 1163,
  \dodoi{10.1086/498708}

\bibitem[{C. {Soubiran} {et~al.}(2010){Soubiran}, {Le Campion}, {Cayrel de
  Strobel}, \& {Caillo}}]{2010AandA...515A.111S}
{Soubiran}, C., {Le Campion}, J.~F., {Cayrel de Strobel}, G., \& {Caillo}, A.
  2010, \bibinfo{title}{{The PASTEL catalogue of stellar parameters},} \aap,
  515, A111, \dodoi{10.1051/0004-6361/201014247}

\bibitem[{R. {Soummer} {et~al.}(2014){Soummer}, {Lajoie}, {Pueyo}, {Hines},
  {Isaacs}, {Nelan}, {Clampin}, \& {Perrin}}]{Soummer2014}
{Soummer}, R., {Lajoie}, C.-P., {Pueyo}, L., {et~al.} 2014, in Society of
  Photo-Optical Instrumentation Engineers (SPIE) Conference Series, Vol. 9143,
  Space Telescopes and Instrumentation 2014: Optical, Infrared, and Millimeter
  Wave, ed. J.~{Oschmann}, Jacobus~M., M.~{Clampin}, G.~G. {Fazio}, \& H.~A.
  {MacEwen}, 91433V, \dodoi{10.1117/12.2057190}

\bibitem[{R. {Soummer} {et~al.}(2012){Soummer}, {Pueyo}, \&
  {Larkin}}]{Soummer2012}
{Soummer}, R., {Pueyo}, L., \& {Larkin}, J. 2012, \bibinfo{title}{{Detection
  and Characterization of Exoplanets and Disks Using Projections on
  Karhunen-Lo{\`e}ve Eigenimages},} \apjl, 755, L28,
  \dodoi{10.1088/2041-8205/755/2/L28}

\bibitem[{T. {Sumi} {et~al.}(2023){Sumi}, {Koshimoto}, {Bennett}, {Rattenbury},
  {Abe}, {Barry}, {Bhattacharya}, {Bond}, {Fujii}, {Fukui}, {Hamada}, {Hirao},
  {Silva}, {Itow}, {Kirikawa}, {Kondo}, {Matsubara}, {Miyazaki}, {Muraki},
  {Olmschenk}, {Ranc}, {Satoh}, {Suzuki}, {Tomoyoshi}, {Tristram}, {Vandorou},
  {Yama}, \& {Yamashita}}]{Sumi2023}
{Sumi}, T., {Koshimoto}, N., {Bennett}, D.~P., {et~al.} 2023,
  \bibinfo{title}{{Free-floating Planet Mass Function from MOA-II 9 yr Survey
  toward the Galactic Bulge},} \aj, 166, 108, \dodoi{10.3847/1538-3881/ace688}

\bibitem[{J. {Teyssandier} {et~al.}(2013){Teyssandier}, {Naoz}, {Lizarraga}, \&
  {Rasio}}]{2013ApJ...779..166T}
{Teyssandier}, J., {Naoz}, S., {Lizarraga}, I., \& {Rasio}, F.~A. 2013,
  \bibinfo{title}{{Extreme Orbital Evolution from Hierarchical Secular Coupling
  of Two Giant Planets},} \apj, 779, 166, \dodoi{10.1088/0004-637X/779/2/166}

\bibitem[{S.~S. {Vogt} {et~al.}(2014){Vogt}, {Radovan}, {Kibrick}, {Butler},
  {Alcott}, {Allen}, {Arriagada}, {Bolte}, {Burt}, {Cabak}, {Chloros},
  {Cowley}, {Deich}, {Dupraw}, {Earthman}, {Epps}, {Faber}, {Fischer}, {Gates},
  {Hilyard}, {Holden}, {Johnston}, {Keiser}, {Kanto}, {Katsuki}, {Laiterman},
  {Lanclos}, {Laughlin}, {Lewis}, {Lockwood}, {Lynam}, {Marcy}, {McLean},
  {Miller}, {Misch}, {Peck}, {Pfister}, {Phillips}, {Rivera}, {Sandford},
  {Saylor}, {Stover}, {Thompson}, {Walp}, {Ward}, {Wareham}, {Wei}, \&
  {Wright}}]{Vogt2014}
{Vogt}, S.~S., {Radovan}, M., {Kibrick}, R., {et~al.} 2014,
  \bibinfo{title}{{APF{\textemdash}The Lick Observatory Automated Planet
  Finder},} \pasp, 126, 359, \dodoi{10.1086/676120}

\bibitem[{H. {von Zeipel}(1910){von Zeipel}}]{vonZeipel1910}
{von Zeipel}, H. 1910, \bibinfo{title}{{Sur l'application des s{\'e}ries de M.
  Lindstedt {\`a} l'{\'e}tude du mouvement des com{\`e}tes p{\'e}riodiques},}
  Astronomische Nachrichten, 183, 345, \dodoi{10.1002/asna.19091832202}

\bibitem[{W. {Vousden} {et~al.}(2021){Vousden}, {Farr}, \&
  {Mandel}}]{2021ascl.soft01006V}
{Vousden}, W., {Farr}, W.~M., \& {Mandel}, I. 2021, \bibinfo{title}{{ptemcee: A
  parallel-tempered version of emcee},} \doeprint{2101.006}

\bibitem[{J.~J. {Wang} {et~al.}(2015){Wang}, {Ruffio}, {De Rosa}, {Aguilar},
  {Wolff}, \& {Pueyo}}]{Wang2015}
{Wang}, J.~J., {Ruffio}, J.-B., {De Rosa}, R.~J., {et~al.} 2015,
  \bibinfo{title}{{pyKLIP: PSF Subtraction for Exoplanets and Disks},},
  Astrophysics Source Code Library, record ascl:1506.001 \doeprint{1506.001}

\bibitem[{J.~J. {Wang} {et~al.}(2016){Wang}, {Graham}, {Pueyo}, {Kalas},
  {Millar-Blanchaer}, {Ruffio}, {De Rosa}, {Ammons}, {Arriaga}, {Bailey},
  {Barman}, {Bulger}, {Burrows}, {Cardwell}, {Chen}, {Chilcote}, {Cotten},
  {Fitzgerald}, {Follette}, {Doyon}, {Duch{\^e}ne}, {Greenbaum}, {Hibon},
  {Hung}, {Ingraham}, {Konopacky}, {Larkin}, {Macintosh}, {Maire}, {Marchis},
  {Marley}, {Marois}, {Metchev}, {Nielsen}, {Oppenheimer}, {Palmer}, {Patel},
  {Patience}, {Perrin}, {Poyneer}, {Rajan}, {Rameau}, {Rantakyr{\"o}},
  {Savransky}, {Sivaramakrishnan}, {Song}, {Soummer}, {Thomas}, {Vasisht},
  {Vega}, {Wallace}, {Ward-Duong}, {Wiktorowicz}, \& {Wolff}}]{Wang2016}
{Wang}, J.~J., {Graham}, J.~R., {Pueyo}, L., {et~al.} 2016,
  \bibinfo{title}{{The Orbit and Transit Prospects for {\ensuremath{\beta}}
  Pictoris b Constrained with One Milliarcsecond Astrometry},} \aj, 152, 97,
  \dodoi{10.3847/0004-6256/152/4/97}

\bibitem[{J.~J. {Wang} {et~al.}(2018){Wang}, {Graham}, {Dawson}, {Fabrycky},
  {De Rosa}, {Pueyo}, {Konopacky}, {Macintosh}, {Marois}, {Chiang}, {Ammons},
  {Arriaga}, {Bailey}, {Barman}, {Bulger}, {Chilcote}, {Cotten}, {Doyon},
  {Duch{\^e}ne}, {Esposito}, {Fitzgerald}, {Follette}, {Gerard}, {Goodsell},
  {Greenbaum}, {Hibon}, {Hung}, {Ingraham}, {Kalas}, {Larkin}, {Maire},
  {Marchis}, {Marley}, {Metchev}, {Millar-Blanchaer}, {Nielsen}, {Oppenheimer},
  {Palmer}, {Patience}, {Perrin}, {Poyneer}, {Rajan}, {Rameau},
  {Rantakyr{\"o}}, {Ruffio}, {Savransky}, {Schneider}, {Sivaramakrishnan},
  {Song}, {Soummer}, {Thomas}, {Wallace}, {Ward-Duong}, {Wiktorowicz}, \&
  {Wolff}}]{Wang2018}
{Wang}, J.~J., {Graham}, J.~R., {Dawson}, R., {et~al.} 2018,
  \bibinfo{title}{{Dynamical Constraints on the HR 8799 Planets with GPI},}
  \aj, 156, 192, \dodoi{10.3847/1538-3881/aae150}

\bibitem[{R.~A. {Wittenmyer} {et~al.}(2007){Wittenmyer}, {Endl}, \&
  {Cochran}}]{2007ApJ...654..625W}
{Wittenmyer}, R.~A., {Endl}, M., \& {Cochran}, W.~D. 2007,
  \bibinfo{title}{{Long-Period Objects in the Extrasolar Planetary Systems 47
  Ursae Majoris and 14 Herculis},} \apj, 654, 625, \dodoi{10.1086/509110}

\bibitem[{N.~F. {Wogan} {et~al.}(2023){Wogan}, {Catling}, {Zahnle}, \&
  {Lupu}}]{Wogan2023}
{Wogan}, N.~F., {Catling}, D.~C., {Zahnle}, K.~J., \& {Lupu}, R. 2023,
  \bibinfo{title}{{Origin-of-life Molecules in the Atmosphere after Big Impacts
  on the Early Earth},} \psj, 4, 169, \dodoi{10.3847/PSJ/aced83}

\bibitem[{Y. {Wu} \& N. {Murray}(2003){Wu} \& {Murray}}]{wu2003}
{Wu}, Y., \& {Murray}, N. 2003, \bibinfo{title}{{Planet Migration and Binary
  Companions: The Case of HD 80606b},} \apj, 589, 605, \dodoi{10.1086/374598}

\bibitem[{J.~W. {Xuan} \& M.~C. {Wyatt}(2020){Xuan} \&
  {Wyatt}}]{2020MNRAS.497.2096X}
{Xuan}, J.~W., \& {Wyatt}, M.~C. 2020, \bibinfo{title}{{Evidence for a high
  mutual inclination between the cold Jupiter and transiting super Earth
  orbiting {\ensuremath{\pi}} Men},} \mnras, 497, 2096,
  \dodoi{10.1093/mnras/staa2033}

\bibitem[{K.~J. {Zahnle} \& M.~S. {Marley}(2014{\natexlab{a}}){Zahnle} \&
  {Marley}}]{2014ApJ...797...41Z}
{Zahnle}, K.~J., \& {Marley}, M.~S. 2014{\natexlab{a}},
  \bibinfo{title}{{Methane, Carbon Monoxide, and Ammonia in Brown Dwarfs and
  Self-Luminous Giant Planets},} \apj, 797, 41,
  \dodoi{10.1088/0004-637X/797/1/41}

\bibitem[{K.~J. {Zahnle} \& M.~S. {Marley}(2014{\natexlab{b}}){Zahnle} \&
  {Marley}}]{Zahnle2014}
{Zahnle}, K.~J., \& {Marley}, M.~S. 2014{\natexlab{b}},
  \bibinfo{title}{{Methane, Carbon Monoxide, and Ammonia in Brown Dwarfs and
  Self-Luminous Giant Planets},} \apj, 797, 41,
  \dodoi{10.1088/0004-637X/797/1/41}

\bibitem[{J. {Zhang} {et~al.}(2024){Zhang}, {Weiss}, {Huber}, {Xuan}, {Bottom},
  {Fulton}, {Isaacson}, {MacDougall}, \& {Saunders}}]{2024arXiv240809614Z}
{Zhang}, J., {Weiss}, L.~M., {Huber}, D., {et~al.} 2024,
  \bibinfo{title}{{Discovery of a Jupiter Analog Misaligned to the Inner
  Planetary System in HD 73344},} arXiv e-prints, arXiv:2408.09614,
  \dodoi{10.48550/arXiv.2408.09614}

\end{thebibliography}
\bibliographystyle{aasjournal}

\end{document}